\documentclass[final,5p,times,twocolumn,authoryear]{elsarticle}

\usepackage{amssymb}
\usepackage{amsmath}
\usepackage[T1]{fontenc}
\usepackage{xspace}
\usepackage{xcolor}

\newcommand{\Damkohler}{Damk{\"o}hler\xspace}
\usepackage{siunitx}

\journal{International Journal of Heat and Fluid Flow}

\begin{document}

\begin{frontmatter}

\title{A model for slip and drag in turbulent flows over superhydrophobic surfaces with surfactant}
\author[inst1]{Samuel D. Tomlinson}
\author[inst2]{Fran\c{c}ois Peaudecerf}
\author[inst3]{Fernando Temprano-Coleto}
\author[inst4]{Frederic Gibou}
\author[inst4]{Paolo Luzzatto-Fegiz}
\author[inst1]{Oliver E. Jensen}
\author[inst1]{Julien R. Landel}
\affiliation[inst1]{organization={Department of Mathematics, University of Manchester},
            addressline={Oxford Rd}, 
            city={Manchester},
            postcode={M13 9PL}, 
            country={UK}}
\affiliation[inst2]{organization={Institute of Environmental Engineering, ETH Z\"{u}rich},
            postcode={8063 Z\"{u}rich}, 
            country={Switzerland}}
\affiliation[inst3]{organization={Andlinger Center for Energy and the Environment, Princeton University},
            city={Princeton},
            postcode={NJ 08544}, 
            country={USA}}
\affiliation[inst4]{organization={Department of Mechanical Engineering, University of California},
            city={Santa Barbara},
            postcode={CA 93106}, 
            country={USA}}
            
\begin{abstract}

Superhydrophobic surfaces (SHSs) can reduce the friction drag in turbulent flows. 
In the laminar regime, it has been shown that trace amounts of surfactant can negate this drag reduction, at times rendering these surfaces no better than solid walls (Peaudecerf et al.,~Proc.~Natl. Acad.~Sci.~USA~114(28), 7254-9, 2017). 
However, surfactant effects on the drag-reducing properties of SHSs have not yet been studied under turbulent flow conditions, where predicting the effects of surfactant in direct numerical simulations remains expensive by today's standards. 
We present a model for turbulent flow inclusive of surfactant, in either a channel or boundary-layer configuration, over long but finite-length streamwise ridges that are periodic in the spanwise direction, with period $P$ and gas fraction $\phi$.
We adopt a technique based on a shifted log law to acquire an expression for the drag reduction.
The average streamwise and spanwise slip lengths are derived by introducing a local laminar model within the viscous sublayer, whereby the effect of surfactant is modelled by modifying the average streamwise and spanwise slip lengths. 
Our model agrees with available laboratory experimental data from the literature when conditions are clean (surfactant-free), or when there are low surfactant levels.
However, we find an appreciable drag increase for larger background surfactant concentrations that are characteristic of turbulent flows over SHSs for marine applications.
\end{abstract}


\end{frontmatter}


\section{Introduction}
\label{sec:Introduction}


Superhydrophobic surfaces (SHSs) combine hydrophobic chemistry and surface roughness to entrap gas layers in their texture, reducing the drag when compared to solid walls. 
Harnessing this feature in turbulent flows could benefit a number of marine, industrial and environmental applications.
For example, SHSs could help reduce energy consumption and associated gas emissions in the shipping industry, which is responsible for around $2.5\%$ of global greenhouse gas emissions and 13\% of NOx and SOx emissions \citep{smith2015third}. 
Early investigations into laminar flows over SHSs modelled the liquid-solid and liquid--gas interfaces as a mixture of no-slip and shear-free boundaries (where the liquid--gas interface is often assumed to be flat), thereby predicting large reductions in drag \citep{Rothstein:2010im}. 
However, recent experimental studies in laminar flow conditions have shown that trace amounts of surfactant can strongly impair the drag-reducing effect of SHSs \citep{Kim2012-iw, Bolognesi2014-vw, peaudecerf17, Song2018-uw}. 
Motivated by these findings, laminar theories have been constructed and compared with numerical simulations inclusive of surfactant \citep{landel19, temprano2021slip}, which demonstrate that surfactant effects should be taken into account to improve model predictions of the drag in channels bounded by SHSs. 

\begin{figure}[t!]
	\centering	\includegraphics[width=\columnwidth]{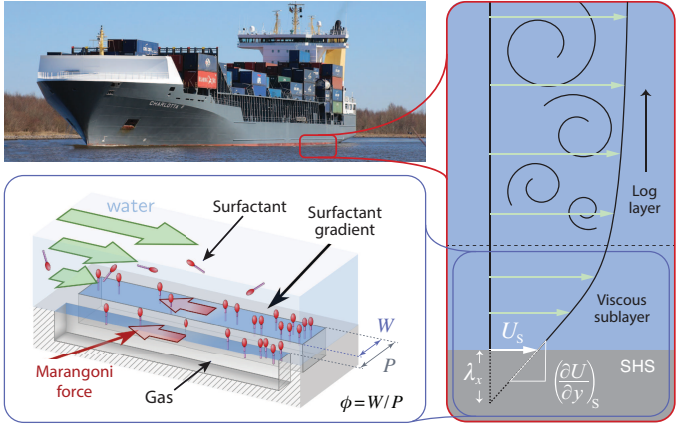}
	\caption{Diagram showing the mechanism by which the presence of surfactant can negatively impact the drag reduction for a flow over a SHS, with period $P$, gas ridge (plastron) width $W$ and gas fraction $\phi = W/P$.
	A buildup of surfactant at the downstream stagnation point of a long but finite-length grating induces an adverse Marangoni force due to the reduction in surface tension \citep{peaudecerf17}.
	The adverse Marangoni force acts to reduce the average streamwise slip length $\lambda_x$ and slip velocity $U_s$ at the interface.
	The smaller average streamwise slip length (or slip velocity) reduces the drag reduction when compared to a surfactant-free flow over a SHS.}
	\label{figure1}
\end{figure}

In this study, we are interested in quantifying the effect of surfactant on the drag reduction in \textit{turbulent} flows over SHSs with long but finite-length streamwise ridges that are periodic in the spanwise direction, for marine applications (see Fig.~\ref{figure1}). 
Surfactant traces have been measured in many natural settings, such as seawater \citep{Pereira2018-nw, frossard2019properties}, rivers, estuaries and fog \citep{Lewis1991-ao, Facchini2000-tu}.
Surfactants can adsorb at liquid--gas interfaces and lower the surface tension between liquid and gas \citep{manikantan2020surfactant}.
They are transported by the flow and accumulate at stagnation points (liquid--gas--solid contact lines), inducing an adverse Marangoni stress at the interface which increases the drag (see Fig.~\ref{figure1}).
In order to model flows inclusive of surfactant, \citet{landel19} assumed that the surfactant concentration is small, and therefore, that there is a uniform interfacial concentration gradient and shear rate along the liquid--gas interface. 
They constructed a scaling theory to model the average streamwise slip and drag in a two-dimensional channel with periodic streamwise ridges in the low-Reynolds-number regime. 
The theory described in \citet{landel19} was extended to three dimensions by \citet{temprano2021slip}. 
In particular, \citet{temprano2021slip}  found that for many small-scale applications, the detrimental effect of surfactants essentially depends on a ratio between a surfactant mobilization length and the grating length. 
The mobilization length depends on the normalised surfactant concentration, Marangoni number, \Damkohler number and Biot number. 
For most small-scale applications, the mobilization length is of the order of centimetres.
If the grating length is larger than the mobilization length, substantial slip, and thus significant drag reduction, can occur, as confirmed by laminar flow experiments \citep{peaudecerf17, Song2018-uw,temprano2021slip}.

Direct numerical simulations (DNSs) that resolve the SHS texture have been used to analyse the mechanisms behind drag reduction in turbulent channel flows with SHS ridges and posts, exclusive of surfactant \citep{Park2013, turk14, jelly14, Rastegari2015, egan2021direct}.
\citet{Park2013} performed DNS to examine the average streamwise slip length and drag in a turbulent channel flow with streamwise grooves that are periodic in the spanwise direction, whilst varying the gas fraction ($\phi \in [0.5, \, 0.94]$) and the ratio of the SHS texture period to the channel height ($P / H \in [0.09, \, 3]$). 
As the period in wall units is increased the viscous sublayer shrinks and the drag reduction appears to converge to the gas fraction of the SHS. 
\citet{turk14} carried out DNS to study the dependence of the drag reduction on the spanwise period of the SHSs in a turbulent channel flow ($P/H \in [0.04, \, 1.56]$).
When the period of the SHS is small, they find that the average streamwise slip length can be predicted by Stokes flow theory \citep{Philip_ZAMP_1972a}; they also show that this approximation breaks down when the period of the SHS becomes larger than approximately twenty wall units.
\citet{Rastegari2015} used DNS to investigate the mechanisms behind turbulent drag reduction for both SHS ridges and posts. 
The drag reduction is decomposed into a gain from the average streamwise slip length and a loss due to modifications to turbulent dynamics and secondary mean flows; these contribute to approximately $80$\% and $20$\% of the total drag reduction, respectively, for the friction Reynolds number of the no-slip flow (${Re}_{\tau_0} = 223$) considered in \citet{Rastegari2015}. 

Experimental studies have investigated the performance of SHSs in internal and external turbulent flows \citep{Daniello_etal_PF_2009, Park2014, xuJFM21}.
\citet{Daniello_etal_PF_2009} found a significant drag reduction in a turbulent channel flow bounded by SHSs with streamwise ridges that are periodic in the spanwise direction, when the viscous sublayer thickness is comparable to the period of the SHS. 
As discussed in \citet{Park2013}, the drag reduction measured by \citet{Daniello_etal_PF_2009} appears to reach a plateau as the viscous sublayer thickness reduces. 
They hypothesised that the drag reduction should asymptote towards the gas fraction, as the viscous sublayer thickness becomes small compared to the SHS texture period.
\citet{Park2014} measured the drag reduction in a turbulent boundary layer flow over a longitudinally ridged SHS test section, which they find increases with increasing gas fraction and period of the SHS. 
However, they did not vary the boundary layer thickness by moving the test section with respect to the upstream origin of the boundary layer or by changing the Reynolds number.
\citet{xuJFM21} investigated the stability of the liquid--gas interface using a towing plate with a SHS test section made of periodic streamwise ridges in open water. 
They measured the drag reduction for varying Reynolds numbers, such that at large Reynolds numbers, they observed that a portion of the upstream region of the SHS grooves became wet. 
They found that reducing the streamwise length of the ridges can improve the drag reduction, due to the enhanced stability of the liquid--gas interface \cite[however, results for laminar flows outlined in][imply that shorter ridges would also make the SHS more susceptible to surfactant effects]{temprano2021slip}.

A review by \citet{Gose2018} of fourteen experimental studies into the turbulent drag reduction for flows over SHSs shows broad discrepancies: the drag reduction ranges from $-90\%$ (i.e. drag increase) to $+90\%$, with five studies finding little ($< 20\%$) or no drag reduction. 
A number of possible causes may explain these discrepancies, as discussed in detail in the review by \citet{park21}.
For example, the liquid--gas interface at the SHS can deform due to pressure differences in the fluid and gas cavity, which has been shown to alter the drag reduction in laminar and turbulent flows over SHSs depending on the protrusion angle \citep{teo09, Rastegari2018-an}.
Alternatively, the turbulence intensity may induce partial or complete wetting of the grooves containing the gas subphase, where the flow would no longer benefit from a flat shear-free interface \citep{rastegari19, xuJFM21}. 
We neglect both of these features of SHSs here for simplicity, and instead focus on the effect of surfactants. 
As previously mentioned, surfactants have been shown to limit the drag-reducing effect of SHSs in laminar flows with a flat liquid--gas interface \citep{peaudecerf17, landel19, temprano2021slip}. 
However, their effect in turbulent flow conditions is yet to be investigated using theory, DNS or experiments.

By exploiting data from DNS which impose average streamwise and spanwise slip lengths at the SHS, semi-empirical models based on a shifted log law have been constructed that predict the drag reduction for turbulent channel flows over SHSs with streamwise ridges that are periodic in the spanwise direction \citep{Fukagata2006-hk, busse12}.
\citet{Fukagata2006-hk} proposed two independent mechanisms that can alter the drag and split the log-law shift into two contributions. 
Their model assumes that the characteristic size of the SHS texture is much smaller than the smallest length scale in the turbulent flow, so that the turbulent flow experiences a spatially averaged slip effect, averaged in planes parallel to the SHS.
The spatially-averaged streamwise slip length increases the mean velocity and decreases the drag. 
The average spanwise slip length decreases the log law velocity and increases the drag. They found that the effect of the spanwise slip length on the drag reduction saturates as the spanwise slip length becomes large, following a nonlinear empirical relationship.
The empirical relationship between the average spanwise slip length and the log law velocity shift proposed by \citet{Fukagata2006-hk} was refined in \citet{busse12}, who performed DNS for flows in SHS channels with streamwise grooves that are periodic in the spanwise direction, where the average slip lengths in the streamwise and spanwise directions are imposed at the boundary. 
Applying the average slip lengths that were imposed as boundary conditions in their DNS to the shifted log law model, both \citet{Fukagata2006-hk} and \citet{busse12} found good agreement between their model and DNS. 
However, neither \citet{Fukagata2006-hk} nor \citet{busse12} related the average streamwise and spanwise slip length to the geometry of the SHS texture, namely the gas fraction and the spanwise period of the SHS, in order to acquire a predictive model that requires only known input parameters.
\citet{luchini2015relevance} related the average slip length to the geometry of the SHS using the laminar solutions due to \citet{Philip_ZAMP_1972a}. 
Luchini's model could provide predictions to  compare with experiments, where the average slip lengths are not known in general and can be hard to measure due to the size of the SHS texture. 
His model predictions for the drag reduction compare well with texture-resolving DNS simulations of turbulent flows over SHS.
However, his predictions agree with DNS results for texture period in wall units up to roughly 30. The poor comparison at larger values may be due to the fact that the log law velocity shift used by \citet{luchini2015relevance} does not saturate, as suggested by the DNS performed by \citet{Fukagata2006-hk} and \citet{busse12}.
Here, we will combine the models proposed by \citet{Fukagata2006-hk}, \citet{busse12} and \citet{luchini2015relevance} to relate the drag reduction to the relevant non-dimensional input parameters related to the flow and  liquid properties and to the geometry, in the case without surfactant. 
Then, we will discuss how this model can be modified to include surfactant effects in order to predict their impact on the drag reduction for turbulent flows over SHS, which is the main objective of our study.

This study investigates the potential effects of surfactant in turbulent flows, for both internal and external geometries, over SHSs made of long but finite-length streamwise ridges that are periodic in the spanwise direction (see Fig.~\ref{figure1}).
We use an existing laminar theory from the literature \citep{landel19} to relate the shear rate at the liquid--gas interface to properties of the fluid, flow, geometry and surfactant. 
This allows us to construct a predictive model that relates the shear rate at the liquid--gas interface to the drag reduction, by combining elements from previous theories \citep{Fukagata2006-hk, busse12, luchini2015relevance}.
We compare our model with available texture-resolving DNS (exclusive of surfactants) and laboratory experimental data in the literature. 
We use our model to discuss the potential role of surfactant in the drag-reduction performance of SHSs for applications in marine transport, where the surfactant concentrations found in natural environments may be much greater than those found in laboratory conditions \citep{Pereira2018-nw, frossard2019properties, temprano2021slip}. 

In Section \ref{sec:Formulation}, we formulate the problem and introduce the quantities used to assess the performance of a SHS: the average streamwise slip length and drag reduction.
In Section \ref{sec:Model}, we formulate a model to assess the performance of a SHS. 
The model is based on the shifted log law for turbulent flow and uses slip lengths that include surfactant effects provided by laminar theories.
In Section \ref{sec:Results}, we present results that compare our model to texture-resolving DNS and laboratory experiments in the literature. 
We then discuss the predictions of our model inclusive of surfactant in relation to the application of drag-reducing SHS in marine environments.
In Section \ref{sec:Conclusions}, we outline key outcomes and extensions of this theory.

\section{Formulation}
\label{sec:Formulation}

\subsection{Superhydrophobic surface flow configuration}
\label{sec:Flow configuration}

\begin{figure}[t!]
	\centering  
    (a) \hfill \hfill \hfill \\[.25cm]
    \includegraphics[width=.9\columnwidth]{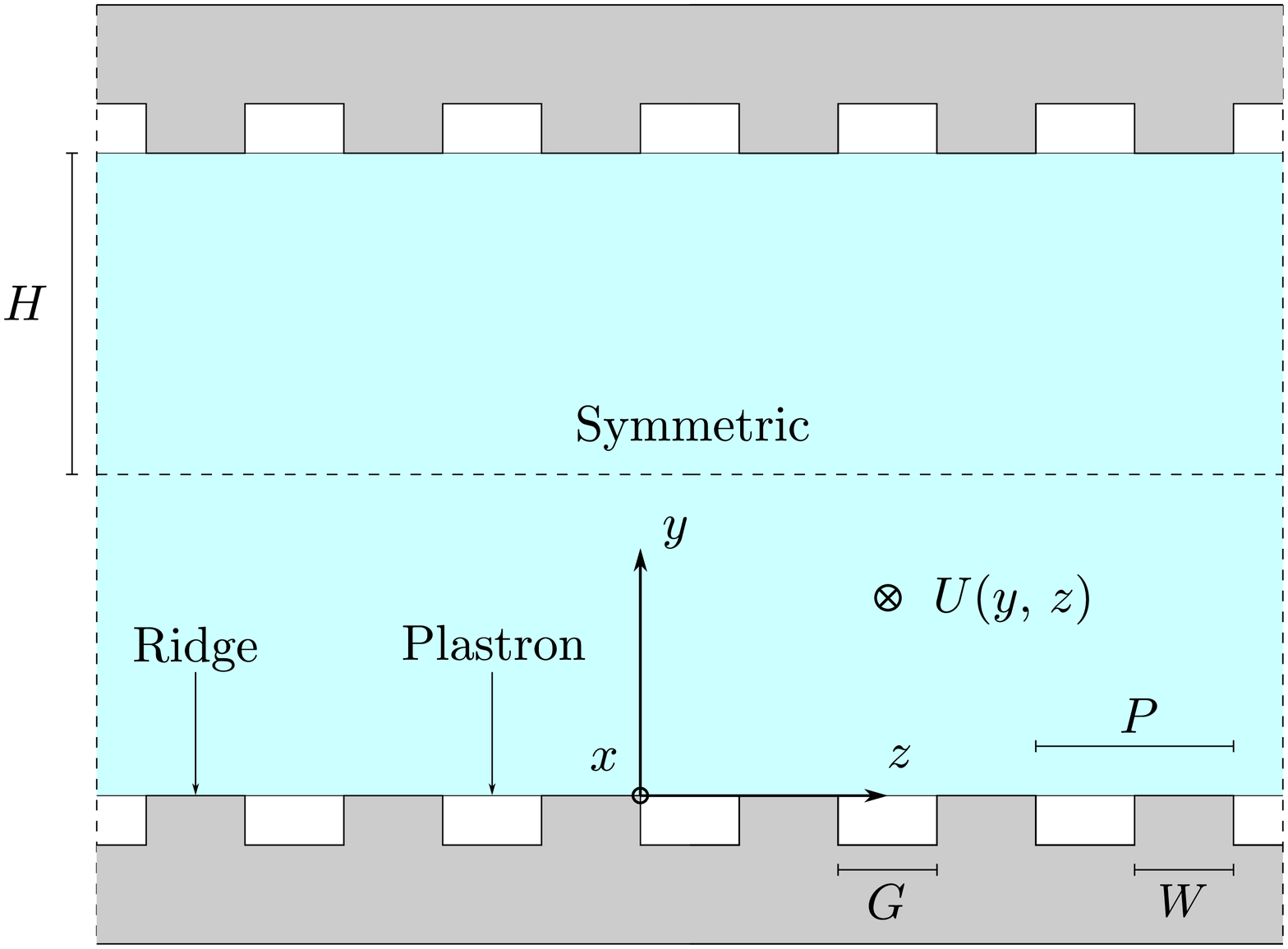}  \\[.25cm]
    (b) \hfill \hfill \hfill \\[.25cm]
    \includegraphics[width=.9\columnwidth]{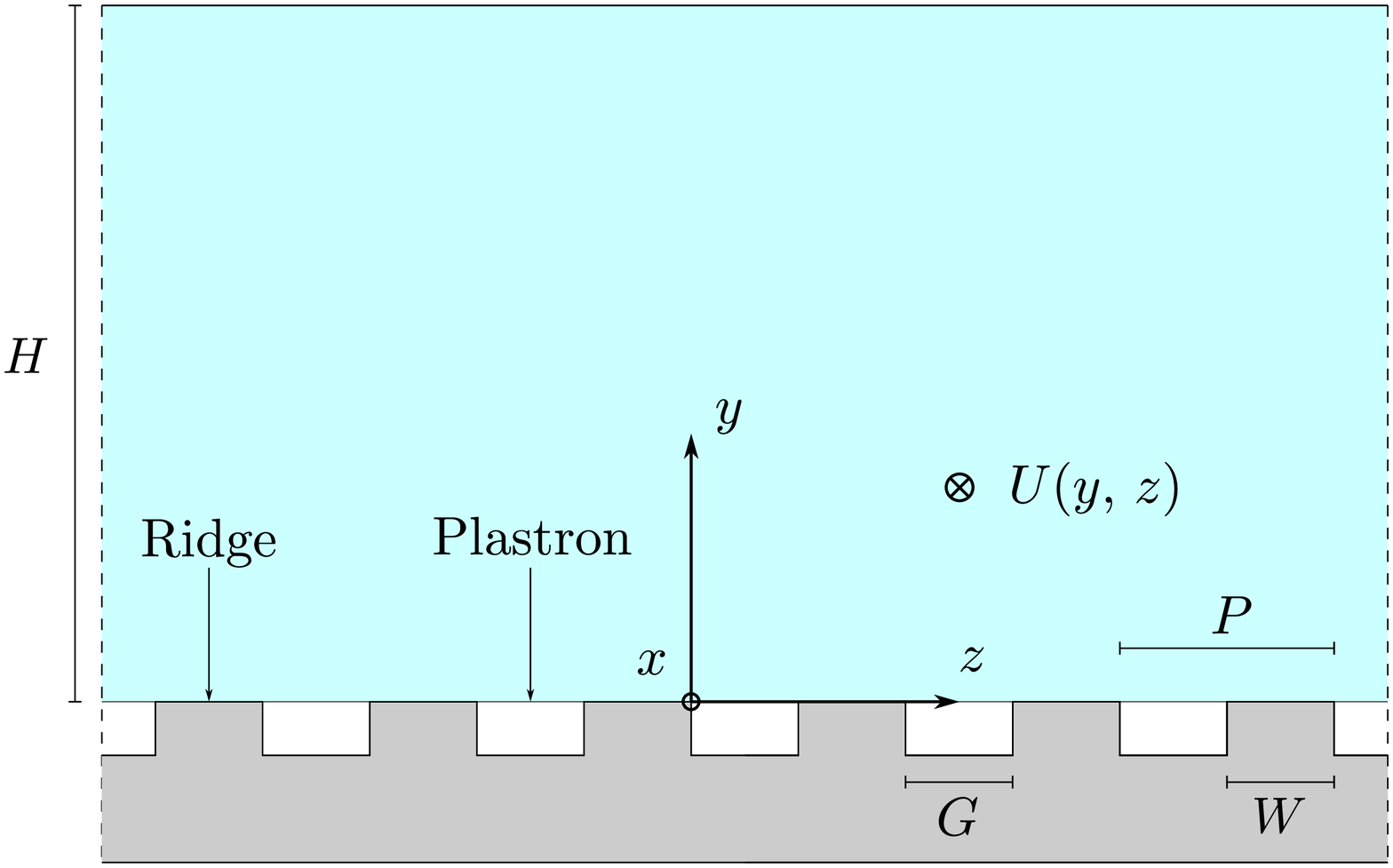}  \\
	\caption{Schematic of the (a) symmetric channel flow with channel height $2H$ and (b) boundary layer flow with boundary layer thickness $H(x)$. 
    The top and bottom walls are made of long but finite superhydrophobic ridges that are periodic in the spanwise direction, such that the liquid is in the Cassie--Baxter state. 
    A shear-rate condition due to the surfactant gradient is assumed at the plastrons and a no-slip condition is assumed at the ridges. 
    The time-averaged fully-developed flow velocity in the streamwise $x$ direction $U$ is assumed invariant with $x$ and periodic in the $z$ direction with period $P$. 
    In this study, we model the drag-reducing effect of the SHS on the flow field, varying Reynolds number, SHS texture geometry and surfactant effects in the turbulent regime. 
    We focus on the periodic flow region for $0\leq z \leq P$, and in the channel flow configuration, we focus on the symmetric region for $0\leq y \leq H$, at any $x$.}
	\label{figure2}
\end{figure}

We consider a channel flow bounded by symmetric SHSs with channel height $2H$ (see Fig.~\ref{figure2}a) and a boundary layer flow over a single SHS with boundary layer thickness $H = H(x)$ (see Fig.~\ref{figure2}b). 
The SHS texture consists of long but finite-length ridges aligned with the main flow direction, where the ridges are periodic in the spanwise direction. 
The liquid is suspended over the SHS texture in the Cassie--Baxter state \citep{Rothstein:2010im}. 
The liquid, assumed incompressible and Newtonian, has dynamic viscosity $\mu$ and density $\rho$. 
A no-slip boundary condition is assumed at the ridge walls.  
We assume that the liquid--gas interfaces (referred to hereafter as `plastrons') are flat, impermeable and have a constant Marangoni shear rate $\gamma_{Ma}$; the Marangoni shear rate is generated by the concentration gradient that arises from surfactant build-up at the downstream stagnation point \citep{landel19}.
We give a description of how \citet{landel19} relate the $\gamma_{Ma}$ to the fluid, flow, geometry and surfactant in \ref{sec:sample:appendix0}.

The three-dimensional time-dependent velocity field is defined by $\boldsymbol{u}=u \boldsymbol{e}_x + v\boldsymbol{e}_y + w\boldsymbol{e}_z$, where $\boldsymbol{e}_x$, $\boldsymbol{e}_y$ and $\boldsymbol{e}_z$ are the unit vectors that describe the streamwise ($x$), wall-normal ($y$) and spanwise ($z$) directions in a Cartesian coordinate frame. 
The origin of the Cartesian coordinate frame is at the bottom SHS, located at $y=0$, on the right-hand-side corner of a ridge at $z=0$. 
A plastron lies at $y=0$ for $0<z<G$, and a ridge lies at $y=0$ for $G<z<P=G+W$, with $G$ the plastron width, $W$ the ridge width and $P$ the period of the SHS texture. 
The velocity vector is decomposed into time-averaged and fluctuating components, assumed to be of the form $\boldsymbol{u} = (U,\,V,\,W)(\boldsymbol{x}) + \boldsymbol{u}'(\boldsymbol{x}, \, t)$, to arrive at the Reynolds-averaged Navier--Stokes equations for a turbulent flow \citep{Pope_CUP_2000}. 
We assume that the streamwise length of the ridges $L$ is finite in order to generate the surfactant gradient that impedes the drag reduction, however, we also assume that $L$ is much larger than $G$, $W$, $P$ and $H$, such that the flow is statistically invariant in the $x$-direction, $U=U(y, \, z)$, and $|V|, \, |W| \ll |U|$. 
In the channel flow configuration only, $U$ is assumed to be symmetric in the $y$-direction with respect to $y=H$.

The friction velocity (or shear velocity) is denoted $U_\tau = \sqrt{\tau/\rho}$ ($U_{\tau_0} = \sqrt{\tau_0/\rho}$ for the no-slip flow), and the viscous length scale is written as $\delta_\tau=\nu/U_\tau$ ($\delta_{\tau_0}=\nu/U_{\tau_0}$ for the no-slip flow), with $\nu=\mu/\rho$ the kinematic viscosity.
Normalizing length scales and velocity scales  using $\delta_\tau$ and $U_{\tau}$ for the SHS flow defines non-dimensional quantities in wall units, which we denote using a superscript $+$. 
To avoid confusion, we typically use the superscript notation with $+$ only for the SHS flow, whereas for the no-slip flow, the normalisation is written explicitly (e.g.~we use $y/\delta_{\tau_0}$ rather than, say, $y^{+0}$).

\subsection{No-slip flow configuration}

As is commonly done in the literature, we compare the SHS flow to a reference flow with conventional no-slip walls, referred to hereafter as the `no-slip flow'. 
More specifically, in the no-slip channel, the SHS texture is replaced by a no-slip wall for all $x$ and $z$. 
Hereafter, we use the subscript $0$ to refer to quantities related to the no-slip flow, which differ from the corresponding quantities for the SHS flow. 
For instance, the time-averaged velocity field in the no-slip flow is $U_0(y)$, which is invariant in both $x$ and $z$.

\subsection{Constant flow rate and constant pressure gradient  conditions}\label{sec:flowconditions}

Two flow conditions have been used in the literature to drive the flow in the SHS and no-slip channels, in order to set up a comparison. 
The SHS and no-slip flows can be driven by imposing the same constant flow rate (CFR), such that the bulk average velocities in both flows are equal and constant, $\overline{U}=\overline{U}_0$.
The overbar $\overline{\cdot}$ represents a spatial average in both the $y$ and $z$ directions. 
Alternatively, the SHS and no-slip flows can be driven by imposing the same constant pressure gradient (CPG), such that the average shear stresses at the boundaries in both flows are equal and constant, $\tau=\tau_0$, where $\tau=\mu\langle \partial U/\partial y \rangle$ at $y=0$ is the time- and space-averaged wall shear stress of the SHS flow, $\tau_0$ is the time-averaged wall shear stress of the no-slip flow, and $\langle \cdot \rangle$ represents a spatial average in the spanwise $z$ direction. 
We include a description of both these conditions here as we convert DNS data from studies performed under CPG conditions to CFR conditions in Section \ref{sec:Results}.

\subsection{Independent non-dimensional parameters}

For the purposes of our study, the SHS flow has four independent non-dimensional parameters, which encode the SHS geometry, surfactant strength and driving condition, whilst the no-slip flow has only one independent non-dimensional parameter, which expresses the driving condition. 
For the SHS flow, two non-dimensional geometric parameters are related to the SHS texture, namely $P/H$ and $\phi$, which express the ratio of the SHS texture period to the wall-normal height and the gas fraction, respectively.
The non-dimensional parameter that represents the surfactant strength, namely $\gamma_{Ma}^+ = \gamma_{Ma}/(\tau/\mu)=\gamma_{Ma}/(U_\tau/\delta_\tau)$, is the time- and space-averaged interfacial shear rate due to surfactant divided by the wall shear rate of the SHS flow.
If the two flows are driven under the CFR condition, the non-dimensional parameters are $Re=  H \overline{U}/ \nu$ and $Re_{0}= H \overline{U}_0/\nu$ which denotes the bulk Reynolds numbers of the SHS and no-slip flows, respectively. 
Under the CFR condition, $Re=Re_0$.
Alternatively, if the two flows are driven under the CPG condition, the remaining  non-dimensional parameters are $Re_\tau=  H U_\tau / \nu$ and $Re_{\tau_0}= H U_{\tau_0} / \nu$, which denotes the friction Reynolds numbers of the SHS and no-slip flows, respectively. 
Under the CPG condition, $Re_\tau=Re_{\tau_0}$. 


\subsection{Superhydrophobic surface performance}
\label{sec:Superhydrophobic surface performance}

There are three main quantities of interest, commonly used in the literature, that characterise the local and global performance of the SHS flow compared to the no-slip flow. 
These quantities are functions of the non-dimensional parameters stated above.

Firstly, the spanwise-averaged streamwise slip length (hereafter designated as the average streamwise slip length) is  defined, dimensionally, as
\begin{equation} \label{eq:sliplengthdefinition}
    \lambda_x = \frac{U_s}{\langle \gamma_I \rangle},
\end{equation}
where $U_s =\langle U_I \rangle$ is the spanwise-averaged slip velocity at the SHS boundary, $U_I(z)$ is the local time-averaged velocity at the SHS boundary $y=0$ (see Fig.~\ref{figure1}) and $\gamma_I(z)=\partial U/\partial y$ is the local time-averaged shear rate at $y=0$. 
The average streamwise slip length $\lambda_x$ represents the extrapolated distance, below the wall, where $U$ vanishes. 
The slip length $\lambda_x$ can be normalised with a relevant length scale, usually either $H$ or $\delta_\tau$, depending on whether the effect of local slip is being compared to the bulk flow, or to the viscous sublayer, respectively.

Secondly, for flows under the CFR condition $\overline{U}=\overline{U}_0$ (i.e. $Re=Re_0$), the drag reduction is defined as
\begin{equation} \label{eq:dragreductiondefinition}
        {DR} = \frac{\tau_0-\tau}{\tau_0} = 1-\frac{Re_\tau^2}{Re_{\tau_0}^2}. 
\end{equation}

Thirdly, for flows under the CPG condition $\tau=\tau_0$ (i.e. $Re_\tau=Re_{\tau_0}$), one defines the added flux, or the relative increase in the bulk-averaged velocity,
\begin{equation}\label{eq:addedfluxdefinition}
        \frac{\Delta \overline{U}}{\overline{U}_0} = \frac{\overline{U}-\overline{U}_0}{\overline{U}_0} = \frac{Re}{Re_0}-1.
\end{equation}

For turbulent flows under the CFR condition, the impact on ${DR}$ of the turbulent flow interactions with the SHS texture can be difficult to interpret for flows near laminar-turbulent transition \citep{turk14}. 
As the friction Reynolds number of the SHS flow is lower than for the no-slip flow (i.e. $Re_\tau<Re_{\tau_0}$), the SHS flow may relaminarize and no longer offer a meaningful comparison to the no-slip flow. 
In contrast, the added flux $\Delta \overline{U}/\overline{U}_0$ compares the SHS and no-slip flow under the CPG condition, such that the friction Reynolds numbers are the same, i.e. $Re_\tau=Re_{\tau_0}$. 
Under the CPG condition, the bulk Reynolds number of the SHS flow increases, i.e. $Re > Re_{0}$, such that a no-slip turbulent flow will correspond to a turbulent SHS flow. 
This increase in bulk Reynolds number tends to have a lesser impact on the global performance of the SHS, as measured through $\Delta \overline{U}/\overline{U}_0$, owing to the homogeneity of the bulk turbulence properties of both SHS and no-slip flows, provided $Re$ and $Re_0$ are both sufficiently large for the  turbulent flows to be fully developed. 
In this study, we assess the global performance of SHSs using ${DR}$ as it is most usually calculated and discussed in experimental studies \citep{Daniello_etal_PF_2009, Park2014, xuJFM21}. 
However, some of the numerical results from the literature \citep{turk14, egan2021direct}, which will be compared to  model predictions, give only $\Delta \overline{U}/\overline{U}_0$, and therefore, we include its discussion and outline a procedure to convert the data to from $\Delta \overline{U}/\overline{U}_0$ to ${DR}$ in \ref{sec:sample:appendix2}.

To evaluate the global performance of a SHS texture, the relationships between ${DR}$ and the relevant independent non-dimensional parameters is sought in the form
\begin{equation}\label{eq:drfeq}
{DR} = f\left({Re}, \, \frac{P}{H}, \, \phi, \, \gamma_{Ma}^+ \right),
\end{equation}
where $Re=Re_0$ under the CFR condition and $f$ is a function to be determined. 
For turbulent flows, $DR$ in \eqref{eq:drfeq} could also be given as a function of $P^+=P/\delta_\tau$, instead of $P/H$ \cite[e.g.][]{Park2013}. 
As mentioned earlier, for Stokes flows and stable laminar flows, the dependence on the Reynolds number can be neglected in \eqref{eq:drfeq} as $Re$ is found to have negligible influence on ${DR}$ \citep{Park2013}. 

\subsection{Reference turbulent no-slip flow model}

For completeness, the canonical turbulent no-slip flow model is reported here. 
A log-law velocity profile holds for $y\gg \delta_{\tau_0}$,
\begin{equation} \label{eq:nosliploglaw}
    \frac{U_0}{U_{\tau_0}} = \frac{1}{\kappa} \ln\left(\frac{y}{\delta_{\tau_0}}\right) + B + \Pi\left( \dfrac{y}{H}\right),
\end{equation}
where $\kappa = 0.41$ is the von K\'{a}rm\'{a}n constant and $B \approx 5.3$ is an empirical constant \citep{Pope_CUP_2000}, and $\Pi$ is the wake function. Note that the net effect of the wake function is expected to be small in our study, as we will discuss in \S\ref{sec:shiftedloglawprofile} when comparing flows with no-slip and SHS boundaries.
In the viscous sublayer ($y\lesssim 10\, \delta_{\tau_0}$), the velocity field of the no-slip flow follows
\begin{equation} \label{eq:nosliplinear}
    \frac{U_0}{U_{\tau_0}} = \frac{y}{\delta_{\tau_0}}.
\end{equation}

The bulk Reynolds number of the no-slip flow is defined as
\begin{equation}
    Re_0 =\frac{\overline{U}_0 H}{\nu}  = \frac{1}{\nu} \int_{y=0}^H U_0 \, \mathrm{d}y.
\end{equation}
The bulk Reynolds number can be found by integrating the velocity profile. 
A common approximation is to neglect the flux associated with the viscous sublayer, thereby integrating the log law from $y=0$ to $H$ \citep{Pope_CUP_2000}. 
To facilitate comparisons with SHS results at relatively low $Re_{\tau_0}$,  we retain the viscous sublayer in the calculation and switch from \eqref{eq:nosliploglaw} to \eqref{eq:nosliplinear} at the value of $y$ for which the two expressions for $U_0$ are equal, which we write as $y = {\beta} \delta_{\tau_0}$, where ${\beta}=(\ln {\beta})/\kappa+B\approx 11.2$ \citep{Pope_CUP_2000}. 
Therefore, $Re_0$ is calculated as
\begin{align} \label{eq:Jintegral}
    Re_0 &= \frac{1}{\delta_{\tau_0}} \int_{y=0}^{{\beta} \delta_{\tau_0}} \frac{y}{\delta_{\tau_0}} \mathrm{d}y  + \frac{1}{\kappa \delta_{\tau_0}} \int_{y={\beta} \delta_{\tau_0}}^H \left[\ln\left(\frac{y}{\delta_{\tau_0}}\right) +\kappa B \right]\mathrm{d}y, \\
    &= \beta \left(\frac{1}{\kappa}- B +\frac{{\beta}}{2} -\frac{\ln ({\beta})}{\kappa} \right) +Re_{\tau_0} \left[ \frac{\ln\left(Re_{\tau_0}\right)}{\kappa} + B -\frac{1}{\kappa}\right]. 
     \label{eq:Jintegral2}
\end{align}
The relative contribution from the first integral in \eqref{eq:Jintegral}, accounting for the viscous sublayer, is usually negligible for no-slip flows (e.g. approximately $0.9\%$ of the total $Re_0$ at $Re_{\tau_0} = 180$). 
However, this term can become significant for SHS flows, where the near-wall fluid can move much faster.

\section{Model}
\label{sec:Model}

\subsection{Low Reynolds number laminar model} \label{subsec:lowreynoldsnumbermodel}

\subsubsection{Laminar slip lengths} 

At low Reynolds numbers, for laminar flows, the slip velocity can be found by solving the incompressible Stokes equation for a linear shear flow in a semi-infinite domain with free-stream shear rate $\tau$. 
At the solid wall, we have no slip.
Following \citet{landel19}, at the liquid--gas interface, the tangential stress balance in the streamwise ($x$) direction can be linearised for small surfactant concentrations, and therefore, we can assume that the surfactant gradient generates a uniform dimensional average Marangoni shear rate denoted by $\gamma_{Ma}$ in the streamwise direction. 
Using transformations detailed in \ref{sec:sample:appendix}, we can solve for the mean streamwise velocity field when $\gamma_{Ma} \neq 0$, building on the solution previously found by \citet{Philip_ZAMP_1972a} for the case $\gamma_{Ma} = 0$.  
The average streamwise slip length including surfactant effects is 
\begin{equation} \label{eq:laminarsliplengthx}
    \lambda_x = \frac{P}{\pi} \left(1 - \frac{\gamma_{Ma}}{\tau/\mu} \right)  \ln \left(\sec\left(\frac{\pi \phi}{2}\right)\right).
\end{equation}
If we define $\gamma_{Ma}^+ = \gamma_{Ma} / (\tau/\mu)$, when $\gamma_{Ma}^+ = 0$ the interface is unaffected by the surfactant (the average streamwise slip length  $\lambda_x$ is maximised) and when $\gamma_{Ma}^+ = 1$ the interface is immobilised by surfactant ($\lambda_x=0$).
However, we leave \eqref{eq:laminarsliplengthx} in terms of $\gamma_{Ma}$ because we will use the laminar scaling theory from \citet{landel19} to relate $\gamma_{Ma}$ to the properties of the flow, geometry, liquid and surfactant, as detailed in  \ref{sec:sample:appendix0}.

If we consider the flow that is perpendicular to the ridges in clean conditions (surfactant-free), the average spanwise slip length is given by \citep{Philip_ZAMP_1972a}
\begin{equation} \label{eq:laminarsliplengthz1}
    \lambda_z = \frac{\lambda_x}{2} = \frac{P}{2 \pi} \ln \left(\sec\left(\frac{\pi \phi}{2}\right)\right) \quad \text{when} \quad \gamma_{Ma}=0.
\end{equation}
However, when surfactants are present, the short spanwise length scale of the SHS implies that the liquid--gas interface are immobilised or close to immobilisation in the spanwise direction as the threshold to achieve immobilisation over short distances is very low \citep{peaudecerf17,temprano2021slip}, such that $\lambda_z = 0$, when $\gamma_{Ma} \neq 0$, i.e. as soon as small amounts of surfactants are present. Therefore, the average spanwise slip length is  given by  
\begin{equation} \label{eq:laminarsliplengthz2}
    \lambda_z = 
    \begin{cases} 
    0 \quad \text{when} \quad \gamma_{Ma} \neq 0, \\[10pt] 
    \displaystyle \frac{P}{2 \pi} \ln \left(\sec\left(\frac{\pi \phi}{2}\right)\right) \quad \text{when} \quad \gamma_{Ma}=0.
    \end{cases}
\end{equation}

\subsubsection{Channel flow configuration}

In order to make a comparison with DNS studies in Section \ref{sec:Results}, we compare the laminar flow in a SHS channel to the no-slip flow in a no-slip channel, in the limit of $H \gg P$. 
In general, the drag reduction can be computed numerically, or using separation of variables and dual series techniques \cite[see e.g.][]{teo09}. 
To compute ${DR}$, one starts from the CFR condition $\overline{U}_0 = \overline{U}$, where
\begin{equation} \label{eq:laminaraverages1}
    \overline{U}_0 = \frac{1}{H P} \int_{y=0}^H \int_{z=0}^P U_0 \, \mathrm{d}y \, \mathrm{d}z,
\end{equation}
and
\begin{subequations} \label{eq:laminaraverages2}
\begin{align}
    \overline{U} = \frac{1}{H P} \int_{y=0}^H \int_{z=0}^P U \, \mathrm{d}y \, \mathrm{d}z.
\end{align}
\end{subequations}
The flow fields, ${U}_0(y)$ and ${U}(y,z)$, are given by the solution to the incompressible Stokes equations. 
The velocity field of the no-slip flow is the canonical Poiseuille solution, leading to
\begin{equation} \label{eq:laminarff1}
    \langle U_0\rangle = {U_0} = \frac{1}{2\mu}\frac{\displaystyle \text{d} p_0}{\text{d} x} y\left(y - 2H\right),
\end{equation}
with $\text{d}p_0/\text{d}x$ the uniform pressure gradient in the no-slip flow. 
In the limit $P/H \ll 1$ for the SHS channel, we can replace the mixed shear-rate/no-slip boundary condition by the homogenised boundary condition $U_s = \langle U_I\rangle = \lambda_x \langle \gamma_I \rangle$, such that the SHS flow has velocity
\begin{equation} \label{eq:laminarff2}
    \langle U\rangle = \frac{1}{2\mu}\frac{\text{d} p}{\text{d} x}\left(y^2 - 2H\left(\lambda_x + y\right)\right),
\end{equation}
with $\text{d}p/\text{d}x$ the uniform pressure gradient in the SHS flow.
Calculating \eqref{eq:laminaraverages1} and \eqref{eq:laminaraverages2} using \eqref{eq:laminarff1} and \eqref{eq:laminarff2}, the bulk average velocities are 
\begin{equation}\label{eq:uBar0Laminar}
    \overline{U}_0 = \frac{1}{H} \int_{y=0}^H \langle U_0\rangle \mathrm{d}y = - \frac{\text{d} p_0}{\text{d} x} \frac{H^2}{3\mu},
\end{equation}
and
\begin{equation}\label{eq:uBarLaminar}
    \overline{U} = \frac{1}{H} \int_{y=0}^H \langle U\rangle \mathrm{d}y = - \frac{\text{d} p}{\text{d} x} \frac{H(H + 3\lambda_x)}{3\mu}.
\end{equation}
Then, using the definition   \eqref{eq:dragreductiondefinition}, the drag reduction can be computed under the CFR condition, $\overline{U}_0 = \overline{U}$, to give $\text{d} p_0/\text{d}x = (3 \lambda_x / H + 1)\text{d} p/\text{d}x$ and
\begin{equation} \label{eq:laminardr0}
    {DR} = \frac{3 \lambda_x}{H + 3\lambda_x}.
\end{equation}


\subsection{Turbulent flow model}

\subsubsection{Shifted log law profile}\label{sec:shiftedloglawprofile}

We assume that the bulk Reynolds numbers $Re$ and $Re_0$ are sufficiently high for the establishment of a fully-developed turbulent flow in both the SHS and no-slip configurations. 
To analyze the effect of surfactants on the drag reduction in the turbulent flow regime, we derive a model based on  the (surfactant-free) shifted log-law technique proposed by \citet{Fukagata2006-hk}, and refined by \citet{busse12} and \citet{luchini2015relevance}. 
The shifted log-law technique is closed using the laminar solutions for the average streamwise ($\lambda_x$) and spanwise  ($\lambda_z$) slip lengths based on semi-infinite shear flows \citep{Philip_ZAMP_1972a}. 
The streamwise and spanwise average slip lengths  can be related to a uniform surfactant-induced Marangoni shear stress $\gamma_{Ma}$ as shown in equations (\ref{eq:laminarsliplengthx}) and (\ref{eq:laminarsliplengthz2}) for $\lambda_x$ and $\lambda_z$, respectively.

Based on classical wall turbulent boundary layer flows, we assume that the turbulent boundary layer flow over the SHS contains two regions of variation close to the SHS boundary: an inner viscous sublayer and an outer log-law layer (see Fig.~\ref{figure1}). 
We assume that $P^+ = P/\delta_\tau \ll 10$.
This assumption implies that the viscous sublayer thickness, of order $10\delta_\tau$, is much larger than the SHS texture period $P$. 
In practice, however, models of this form provide reasonable approximations up to $P^+ \lessapprox 25$ \citep{fairhall19}. 
The flow near the SHS is  homogenised by viscosity within the viscous sublayer since the layer affected by the SHS texture has a thickness of order $P$ \citep{Philip_ZAMP_1972a, Ybert2007-mi}. 
Thus, the SHS texture affects the turbulent bulk flow via homogenised quantities, such as the average streamwise and spanwise slip lengths.

In the outer region, corresponding to $y^+\gg 1$, the bulk flow velocity over the SHS is assumed to follow the shifted log-law model \citep{Fukagata2006-hk}
\begin{equation} \label{eq:sliploglaw}
    U^+ = \frac{1}{\kappa} \ln\left(y^+\right) + B + \Delta U^+(\lambda^+_x, \lambda^+_z),
\end{equation}
where $U^+=U/U_\tau$. 
For the boundary layer flows considered herein, the log laws (\ref{eq:nosliploglaw}, \ref{eq:sliploglaw}) could be extended to include a wake function \citep{Pope_CUP_2000}.
However, if we assume that the wake function is the same over both a SHS and solid wall, then these terms will have a small effect on the drag reduction calculation.
The term $\Delta U^+$ is modelled as \citep{busse12}
\begin{equation} \label{eq:shift}
\Delta U^+(\lambda^+_x, \lambda^+_z) = U_s^+ - \Delta U_{\text{loss}}^+ = \lambda_x^+ - \frac{4 \lambda_z^+}{4 + \lambda_z^+}.
\end{equation}
In \eqref{eq:shift}, $U_s^+$  describes the gain (positive shift in $U^+$) due to the streamwise slip length, since $U_s^+=\lambda^+_x$ in wall units by definition. 
The term $\Delta U_{\text{loss}}^+$ reflects the losses (negative shift in $U^+$) due to spanwise turbulent momentum transfer. 
The quantity $\Delta U_{\text{loss}}^+$ is related to the normalised spanwise slip length, $\lambda^+_z$, through the empirical relationship proposed by \citet{busse12}, that is $\Delta U_{\text{loss}}^+ =  4 \lambda_z^+/(4 + \lambda_z^+)$.  
An alternative relationship for $\Delta U_{\text{loss}}^+$ in \eqref{eq:shift} was proposed by \citet{Fukagata2006-hk}, based on an exponential dependence with $\lambda_z^+$. 
We choose to employ the relationship of \citet{busse12} because of its simplicity and accuracy. 
We note that the modelling approach above is inspired by the work of \citet{luchini91} on riblets. 
Riblets are another type of passive drag-reducing surface using geometrical surface undulations at the boundary, which can modify the turbulent flow near the boundary to reduce drag.  
\citet{luchini91}  proposed that for riblets $\Delta U^+= \lambda_x^+ - \lambda_z^+$, which is the linearised form of (\ref{eq:shift}) and does not account for the saturation effects later proposed by \citet{Fukagata2006-hk} and \citet{busse12} for SHSs \cite[see][for a recent review on riblets and SHSs]{ibrahim21}.

\subsubsection{Average slip lengths}

The model in \eqref{eq:sliploglaw} must be closed to provide a fully predictive relationship, in the form \eqref{eq:drfeq}, for the drag reduction $DR$ as a function of the relevant input non-dimensional parameters: $Re$ the bulk Reynolds number, $P/H$ or $P^+$ the non-dimensional texture period, $\phi$ the gas fraction, and $\gamma_{Ma}^+$ the non-dimensional Marangoni shear rate due to the effect of surfactant, which could be set to zero for  surfactant-free flows. 

We close the model in \eqref{eq:sliploglaw} and \eqref{eq:shift} following the approach proposed  by \citet{luchini91} for riblets \cite[see also][for SHS]{luchini2015relevance}.
Since the flow in the viscous sublayer is dominated by viscosity, we assume that the average streamwise and spanwise slip lengths $\lambda_x^+$ and $\lambda_z^+$ follow  the Stokes flow solutions \eqref{eq:laminarsliplengthx} and \eqref{eq:laminarsliplengthz2} (normalised in wall units), which provide the dependence on $P/H=P^+/Re_\tau$, $\phi$ and $\gamma_{Ma}^+$. 
We couple the flow within the viscous sublayer with the turbulent flow in the log-layer \eqref{eq:sliploglaw} through the characteristic shear 
 rate driving the Stokes flow problems leading to $\lambda_x^+$ and $\lambda_z^+$ in \eqref{eq:laminarsliplengthx} and \eqref{eq:laminarsliplengthz2}. 
The shear rate $\tau$ can then be related to $Re$ by integrating the velocity profile \eqref{eq:sliploglaw}, as shown in the next section, thereby fully closing the model for $DR$.

The normalisation of the average streamwise slip length in wall coordinates is well defined through $\lambda_x^+=\lambda_x/\delta_\tau$.
However, the normalisation of the average spanwise slip length in wall coordinates, $\lambda_z^+$, is more subtle \citep{turk14, Seo2016-dw}. 
Since the average shear stress in the spanwise direction is zero, by definition in this problem, it is unclear what the imposed stress should be for the spanwise Stokes flow leading to \eqref{eq:laminarsliplengthz1}, and thus how $\lambda_z$ should be normalised. 
To resolve this uncertainty, we note that $\Delta U_{\text{loss}}^+$ in \eqref{eq:shift} represents the homogenised effect of the spanwise turbulent momentum transfer related to the turbulent flow interactions and the SHS texture through the viscous sublayer. 
We assume that the spanwise velocity fluctuations at the origin of the spanwise turbulent momentum transfer scale with the streamwise velocity fluctuations. 
This assumption is commonly made for wall turbulent boundary layers \citep{Pope_CUP_2000}.
It implies that the outer flow is homogenised in such a way that the average and fluctuating bulk shear stress in the streamwise and spanwise directions are of the same order of magnitude as the prescribed streamwise shear stress $\tau$ (the only characteristic shear stress in the problem). 
Therefore, we normalise both the streamwise and spanwise average slip lengths using $\delta_\tau$, with $\lambda_x$ from \eqref{eq:laminarsliplengthx} and $\lambda_z$ from \eqref{eq:laminarsliplengthz2}. 

\subsubsection{Drag reduction}\label{sec:DRVHR}

To compute ${DR}$ and determine the relationship with known input parameters \eqref{eq:drfeq}, we impose the CFR condition, $\overline{U}_0 =  \overline{U}$, or equivalently $Re_0=Re$, with
\begin{equation}
   Re = \frac{1}{ \nu} \int_{y=0}^H   \langle U\rangle \, \mathrm{d}y.
\end{equation}
We decompose the SHS flow between the outer turbulent bulk flow, which follows the shifted log law \eqref{eq:sliploglaw} and with bulk Reynolds number $Re_\text{log}$, and the flow in the inner viscous sublayer, which we approximate by the Stokes solution described in Section \ref{subsec:lowreynoldsnumbermodel} and with bulk Reynolds number $Re_\text{sub}$, such that $Re = Re_\text{log} + Re_\text{sub}$, where
\begin{align}
    Re_\text{log} &= \frac{1}{\nu} \int_{y={\beta} \delta_{\tau}}^H U \mathrm{d}y \\
    &= \frac{{\beta} }{\kappa } \left\{1-\kappa \left[ B + \Delta U^+(\lambda_x^+,\lambda_z^+)\right] -\ln ({\beta}) \right\} \nonumber \\
    & \quad \quad +\frac{Re_{\tau}}{\kappa} \left\{ \ln\left(Re_{\tau}\right) + \kappa \left[ B + \Delta U^+(\lambda_x^+,\lambda_z^+)\right] -1\right\}, \label{eq:K1integral}
\end{align}
with $\Delta U^+(\lambda_x^+,\lambda_z^+)$ given in \eqref{eq:shift}, and $\lambda_x$ and $\lambda_z$ given in \eqref{eq:laminarsliplengthx} and \eqref{eq:laminarsliplengthz1}--\eqref{eq:laminarsliplengthz2}, respectively; and 
\begin{equation} \label{eq:K2integral}
    Re_\text{sub} = \frac{1}{P \delta_{\tau}} \int_{z=0}^{P} \int_{y=0}^{{\beta} \delta_{\tau}} U_P^+  \, \mathrm{d}y \, \mathrm{d}z.
\end{equation}
The velocity field $U_P$ inside the viscous sublayer is given by (see \ref{sec:sample:appendix}),
\begin{equation} \label{eq:philip}
    U_P^+ = y^+ + \Im\left(\frac{P^+}{\pi}\left(1 - \gamma_{Ma}^+ \right) \arccos\left(\frac{\cos\left(\displaystyle \frac{\pi \theta^+}{P^+}\right)}{ \cos\left(\displaystyle\frac{\pi \phi}{2} \right)}\right) -  \theta^+\right),
\end{equation}
where $\theta = z + i y$, $i^2 = -1$ and $\Im(\cdot)$ denotes the imaginary part. 
Combining these with the CFR condition $Re_0=Re$, we have an implicit equation relating $Re_\tau$ and $Re_{\tau_0}$, as well as all the other relevant  non-dimensional parameters  $Re$, $P^+$, $\phi$ and $\gamma_{Ma}^+$,
\begin{multline} \label{eq:equivilentflux2highre}
    Re_0 \left(Re_{\tau_0}, \, P^+,\, \phi, \, \gamma_{Ma}^+ \right) = Re_\text{log} \left(Re_{\tau},\, P^+,\, \phi,\, \gamma_{Ma}^+ \right) \\ + Re_\text{sub} \left(Re_{\tau},\, P^+,\, \phi,\, \gamma_{Ma}^+ \right),
\end{multline}
where $Re_0$ is given by \eqref{eq:Jintegral}, $Re_\text{log}$ by \eqref{eq:K1integral} and $Re_\text{sub}$ by \eqref{eq:K2integral}. 
Incidentally, the contribution from $Re_\text{sub}$ can often be more than 5\% of the total. 
We solve \eqref{eq:equivilentflux2highre} numerically to compute the ratio $Re_\tau/Re_{\tau_0}$ and calculate ${DR}=1-(Re_\tau/Re_{\tau_0})^2$ according to \eqref{eq:dragreductiondefinition}, as a function of $Re=Re_0$, $P^+$, $\phi$ and $\gamma_{Ma}^+$.



\section{Results}
\label{sec:Results}

\subsection{Comparison with direct numerical simulations excluding surfactant} \label{ref:Comparison with direct numerical simulations excluding surfactant}

\subsubsection{Average streamwise slip length and drag reduction}

\begin{figure}[t!]
	\centering
    (a) \hfill \hfill \hfill \\[.25cm]
    \includegraphics[width=\columnwidth,trim={3cm 0 4cm 1.5cm},clip]{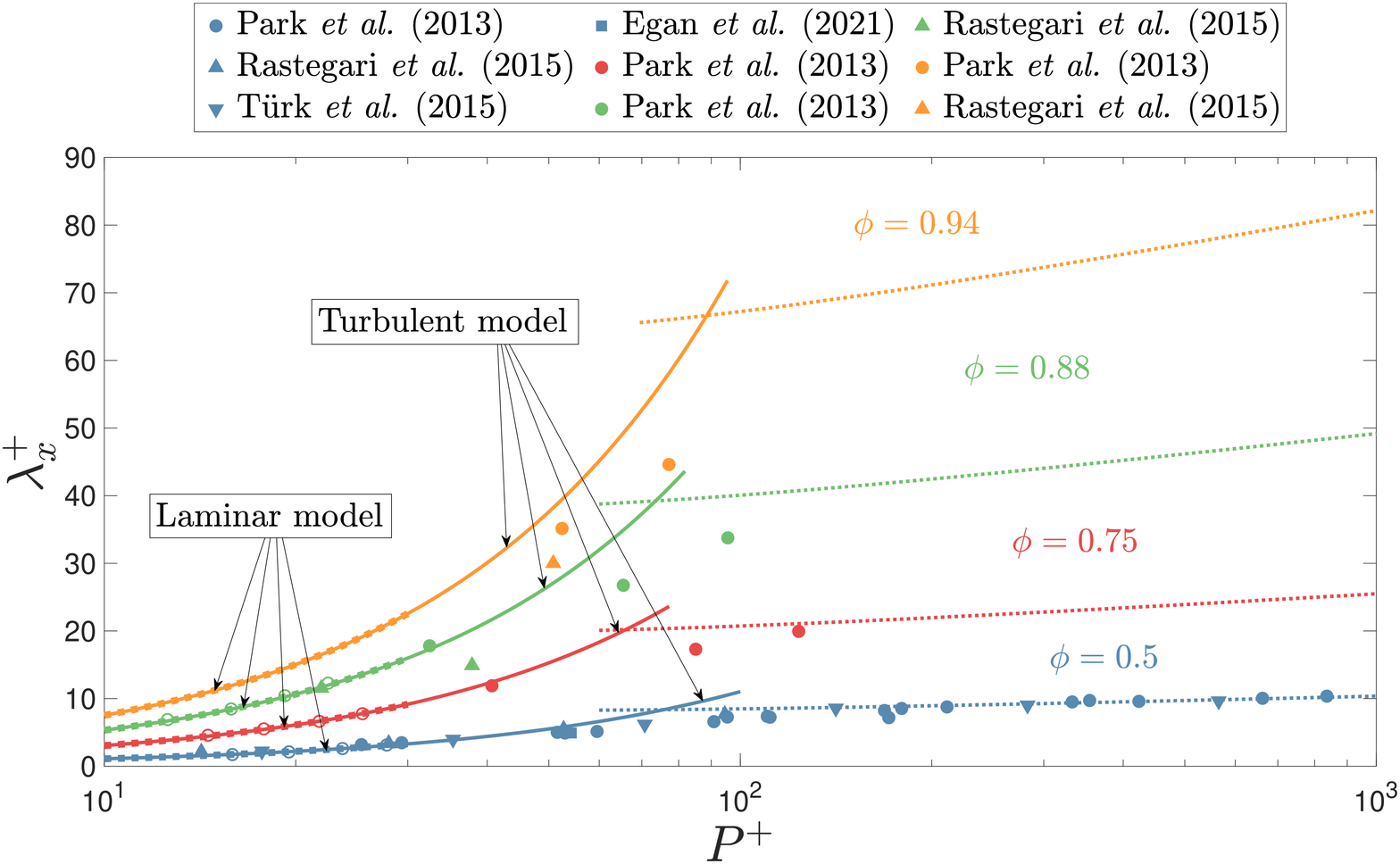} \\
    (b) \hfill \hfill \hfill \\[.25cm]
    \includegraphics[width=\columnwidth,trim={3cm 0cm 4cm 6cm},clip]{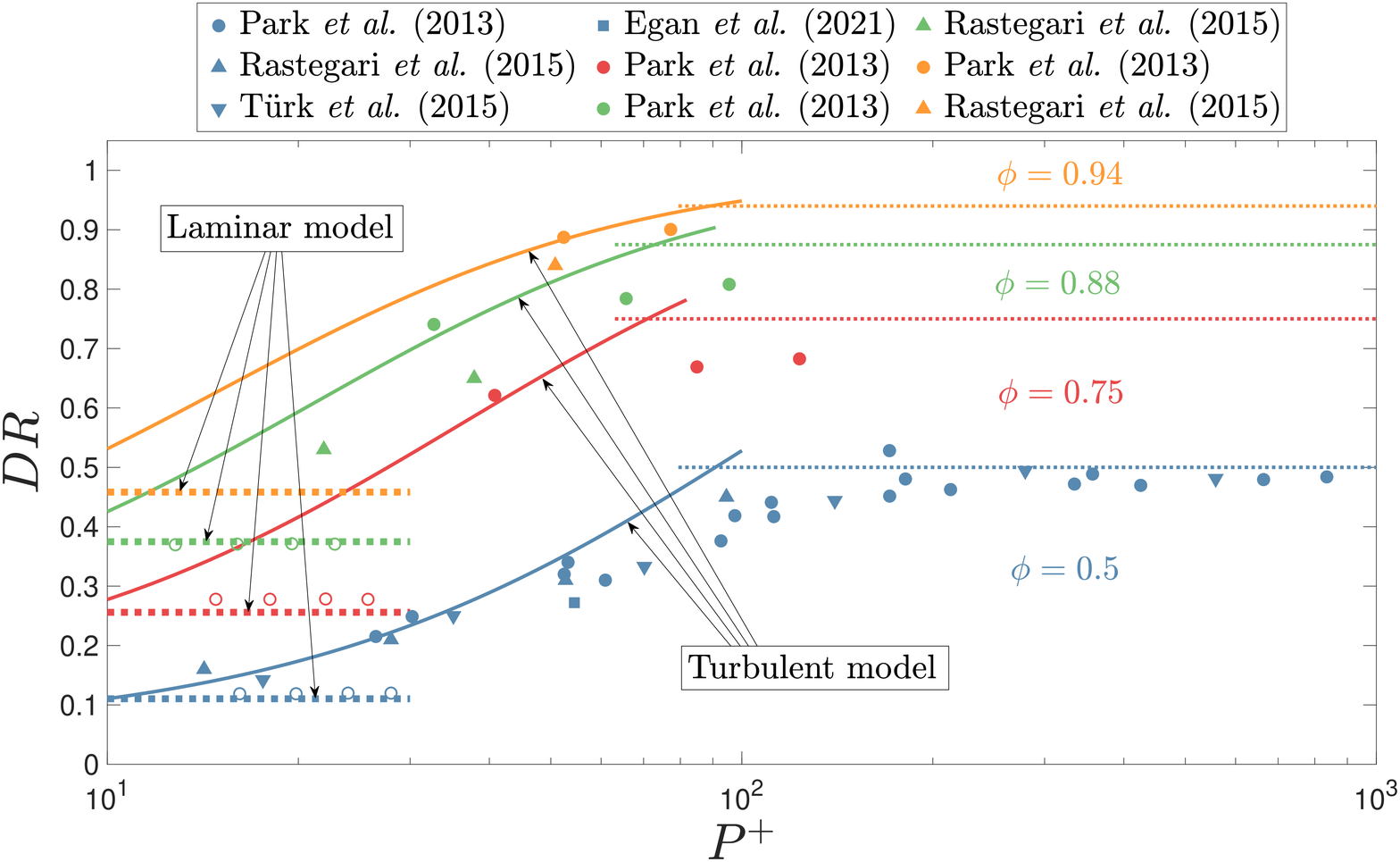}
	\caption{Comparison of our laminar and turbulent model predictions for (a) the average streamwise slip length in wall units $\lambda_x^+$ using \eqref{eq:laminarsliplengthx} and (b) the drag reduction ${DR}$ using (\ref{eq:laminardr0}) (laminar model) and (\ref{eq:dragreductiondefinition}, \ref{eq:equivilentflux2highre}) (turbulent model) with texture-resolving DNS data in the literature \citep{Park2013, turk14, Rastegari2015, egan2021direct}, whilst varying the period in wall units ($P^+$) for different gas fractions ($\phi$), where the average Marangoni shear rate $\gamma_{Ma}=0$. 
	Open symbols and dashed lines represent laminar simulations and theory ($P^+ \in [10, \, 30]$), filled symbols and solid lines represent turbulent simulations and theory ($P^+ \in [10, \, 100]$), and dotted lines show  ${DR} = \phi$ in (b) and the corresponding turbulent predictions for $\lambda_x^+$ using this $\phi$ in (a) ($P^+ \in [100, \, 1000]$).}
	\label{figure3}
\end{figure}

In Fig.~\ref{figure3}, we compare our laminar and turbulent model predictions (excluding surfactant effects, such that the average Marangoni shear rate $\gamma_{Ma} =0$) with available texture-resolving DNS (also exclusive of surfactant) for turbulent channel flows bounded by SHSs with long streamwise ridges that are periodic in the spanwise direction \citep{Park2013, turk14, Rastegari2015, egan2021direct, park21}, as a function of the SHS texture period in wall units $P^+ \in [0, \, 100]$ and gas fraction $\phi = 0.5$ (blue symbols and lines), $\phi = 0.75$ (red), $\phi = 0.88$ (green) and $\phi = 0.94$ (yellow). 
We investigate two quantities that are commonly used to characterise the local and global performance of SHSs compared to the no-slip flow: in Fig.~\ref{figure3}(a) we show the average streamwise slip length in wall units $\lambda_x^+$ and in Fig.~\ref{figure3}(b) we show the drag reduction ${DR}$.

We first comment briefly on the regime transition that takes place between the laminar flow regime (DNS data and theory shown with open symbols and dashed lines, respectively, plotted for $P^+ \in [10, \, 30]$) and the turbulent flow regime (DNS data and theory shown with filled symbols and solid lines, respectively, plotted for $P^+ \in [10, \, 100]$), which has been discussed previously \citep{Rothstein:2010im, Martell2010, Park2013, Rastegari2015, Seo2016-dw, park21}.
In Fig.~\ref{figure3}(a), the theoretical predictions for the average streamwise slip length $\lambda_x^+$ in wall units are the same in both the laminar and turbulent models following \eqref{eq:laminarsliplengthx}. 
The theoretical predictions are in excellent agreement with the DNS data performed by \citet{Park2013} in both the laminar and turbulent flow regimes. 
This confirms the modelling assumption that the average streamwise slip length $\lambda_x$ for turbulent flows over SHS can be modelled using a Stokes flow model, as also found by \citet{turk14}.
In Fig.~\ref{figure3}(b), the drag reduction predicted by the laminar model using \eqref{eq:laminardr0} does not vary with $P^+$, as expected since for laminar flows the drag reduction does not depend on the Reynolds number. 
In contrast, the drag reduction predicted by the turbulent model using (\ref{eq:dragreductiondefinition}, \ref{eq:equivilentflux2highre}) increases rapidly with $P^+$, also in agreement with the turbulent DNS data  by \citet{Park2013, turk14, Rastegari2015, egan2021direct}.
This change in drag-reduction behaviour, from laminar to turbulent flow, is associated with the development of a turbulent boundary layer near the SHS boundary, where the viscous sublayer thickness ($\sim 10\delta_\tau$) replaces the channel height $H$ as the relevant length scale when evaluating the drag reduction \citep{Rothstein:2010im}. 
Due to the limited amount of turbulent DNS data for $\phi > 0.5$,  the dependence of the laminar--turbulent transition on the gas fraction is not entirely clear. 
Nevertheless, the DNS data suggest a similar transition at all gas fractions studied, which is captured by the laminar and turbulent model predictions at different gas fractions (shown with different colours)
In Fig.~\ref{figure3}, both the average streamwise slip length in wall units (panel a) and the drag reduction (panel b) increase as the gas fraction of the SHS increases. 

Next, we comment on the other regime transition that takes place at large $P^+$ as the turbulent DNS data change trend, which has been observed and discussed previously \citep{Rothstein:2010im, Park2013, turk14, seo15, Seo2018, fairhall19, rastegari19, park21}. 
This change in trend also corresponds to a departure from the turbulent theory.  As a matter of fact, turbulent model predictions (solid lines) and the DNS data (solid symbols) agree well  for both $\lambda_x^+$ and ${DR}$ for $P^+=P/\delta_\tau \lessapprox 50$ at all gas fractions (Fig.~\ref{figure3}). 
In this regime,  the viscous sublayer thickness ($\sim 10\delta_\tau$) is large or comparable to the SHS period $P$, which effectively corresponds to, and even extends, the regime of validity of the model, which was assumed to be valid for $P^+\ll 10$. 
However, for $P^+ \gtrapprox 50$, $\lambda_x^+$ and ${DR}$ increase more slowly with $P^+$ than the theoretical predictions; as found in \citet{Park2013}. 
This behaviour is in contrast with the turbulent model, which predicts an increase to $100\%$ drag reduction for $P^+ \rightarrow \infty$. 
However, the departure of the model from the DNS data at large $P^+$ is not unexpected as the regime of validity of our turbulent model clearly breaks down as $P^+\gg 10$. 
The DNS data strongly suggest that the drag reduction saturates for increasing $P^+\gg 10$ \citep{Park2013, turk14, Rastegari2015, egan2021direct}.
In the limit of $P^+\to \infty$, we expect that the drag reduction asymptotes towards the gas fraction of the SHS (dotted lines), as also discussed by \citet{Daniello_etal_PF_2009, Rothstein:2010im, park21}.
Using this empirical asymptote, for a given $P^+$, we can calculate the $\lambda_x^+$ that gives ${DR} = \phi$ using the turbulent model (\ref{eq:dragreductiondefinition}, \ref{eq:equivilentflux2highre}) (see the dotted lines in Fig.~\ref{figure3}a).
We can improve the model predictions for applications at large $P^+$ by taking the streamwise slip length in wall units $\lambda_x^+$ and drag reduction ${DR}$ to be the value predicted using the turbulent models, \eqref{eq:laminarsliplengthx} and (\ref{eq:dragreductiondefinition}, \ref{eq:equivilentflux2highre}) respectively, for $P^+ \lessapprox 70$, and the values corresponding to the empirical asymptote ${DR} = \phi$ for $P^+ \gtrapprox 70$. 
For $P^+ \approx 70$, a transition region appears, as noticeable in particular for $DR$ (see Fig.~\ref{figure3}b), which could be modelled using a composite function of the turbulent model and the gas fraction.

\subsubsection{Average streamwise velocity field}

\begin{figure}[t!]
	\centering
    (a) \hspace{3.75cm} (b) \hfill \hfill \hfill \\[.1cm]  \includegraphics[width=.47\columnwidth,trim={0cm 0 1cm 1cm},clip]{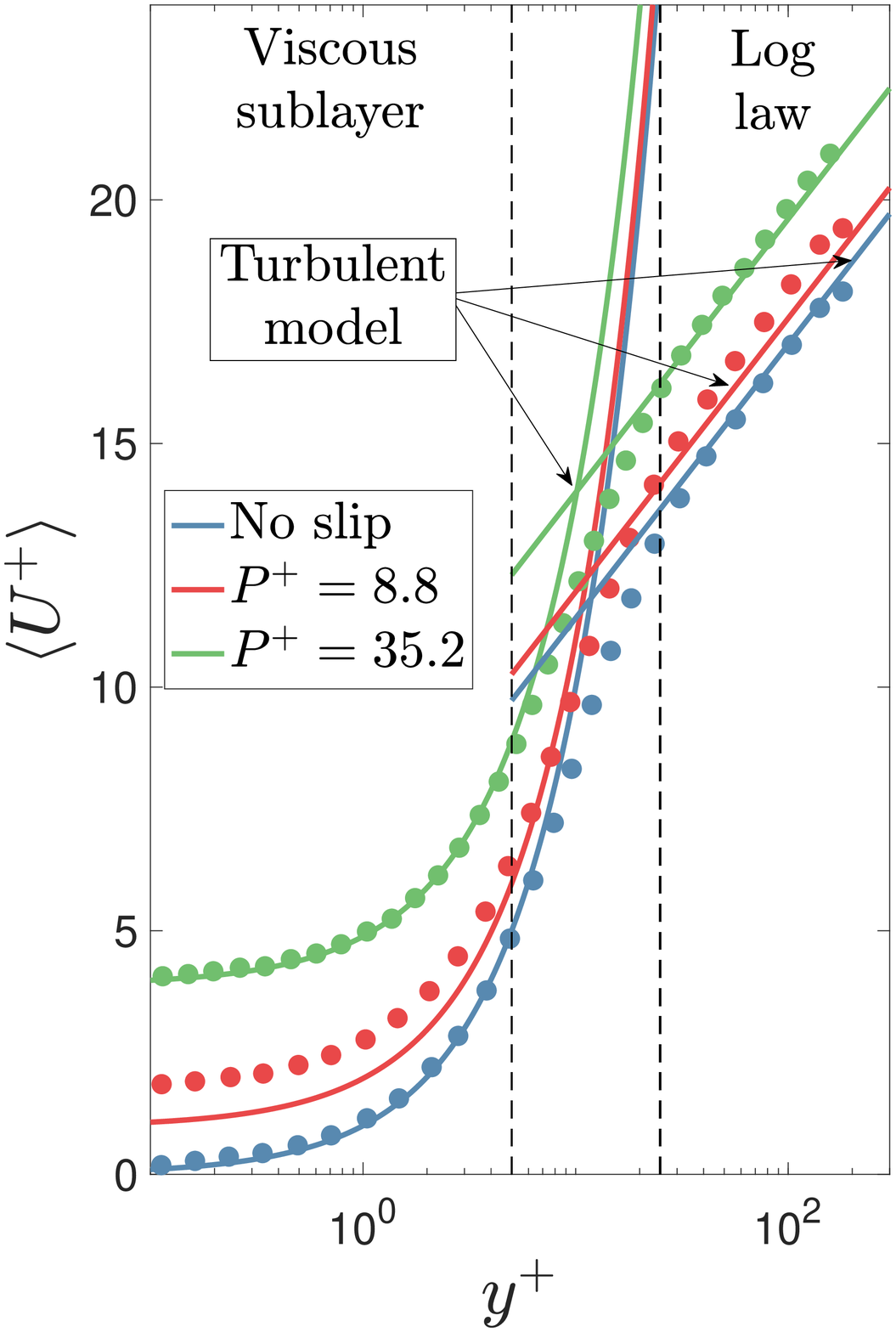} \hfill \includegraphics[width=.47\columnwidth,trim={0cm 0cm 1cm 1cm},clip]{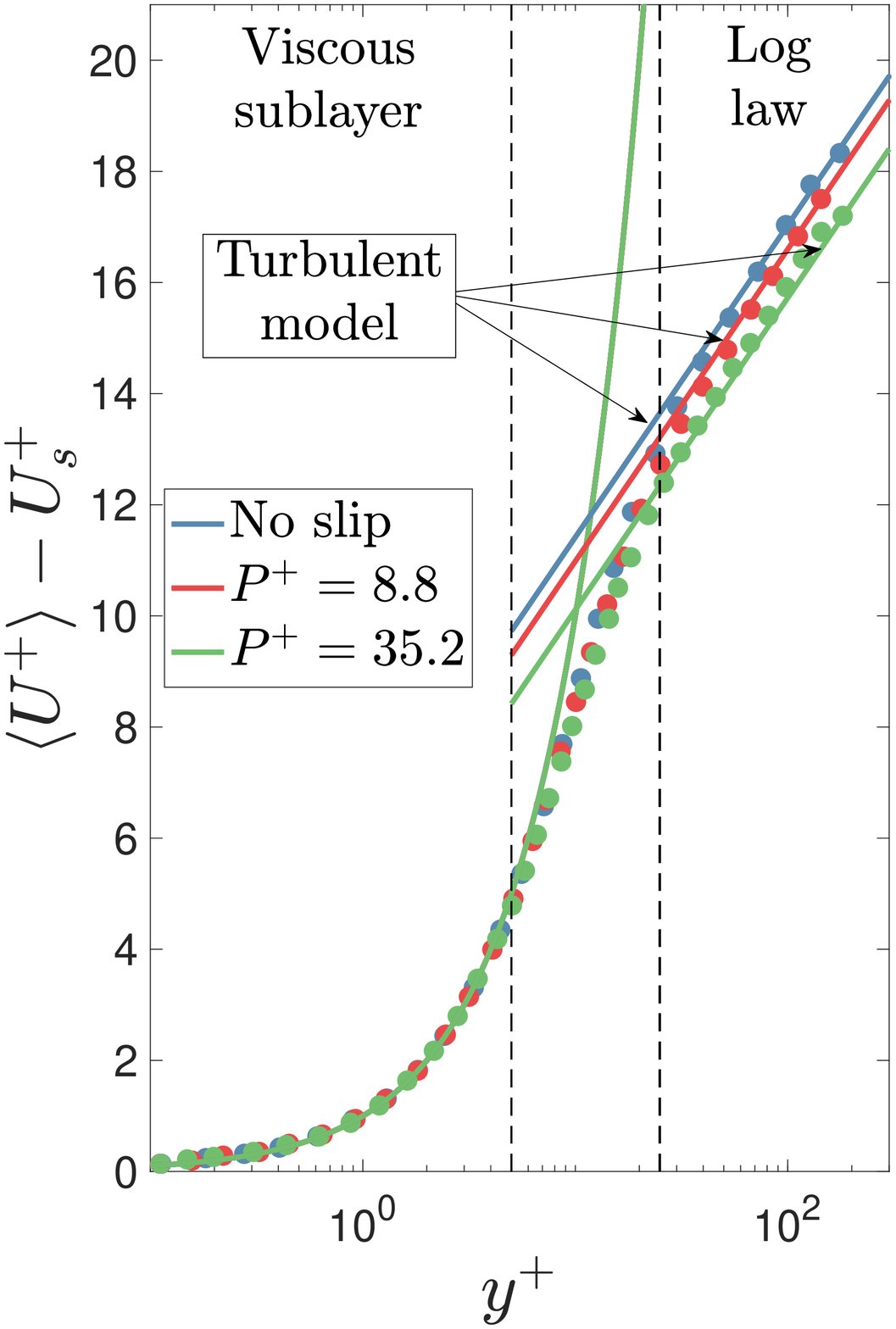} \hfill \hfill \hfill \\
	\caption{Comparison of our turbulent model predictions (solid lines) using (\ref{eq:sliploglaw}, \ref{eq:shift}, \ref{eq:philip}) for (a) the average streamwise velocity in wall units ($\langle U^+\rangle$) and (b) the average streamwise velocity shifted by the slip velocity in wall units ($\langle U^+\rangle - U_s^+$) with texture-resolving DNS data in the literature \citep{turk14} (symbols), whilst varying the wall-normal distance from the SHS in wall units ($y^+$), for different period lengths in wall units ($P^+$) and gas fraction $\phi=0.5$, where the average Marangoni shear rate $\gamma_{Ma}=0$.}
	\label{figure4}
\end{figure}

In Fig.~\ref{figure4} we compare our turbulent model predictions for the average streamwise velocity profile $\langle U^+ \rangle$ as a function of $y^+>0$ across the viscous sublayer and  log law regions with available texture-resolving DNS data from  \cite{turk14}. 
Both the turbulent model and DNS data exclude surfactant effects, i.e. $\gamma_{Ma} = 0$.
As a reference, we show the canonical no-slip flow ($\phi=0$) with blue symbols (DNS data) and blue solid lines (model) using \eqref{eq:nosliploglaw}. 
The other profiles are at gas fraction $\phi = 0.5$ for $P^+ = 8.8$ (red) and $P^+ = 35.2$ (green) using (\ref{eq:sliploglaw}, \ref{eq:shift}, \ref{eq:philip}).
We examine the average streamwise velocity profile in wall units $\langle U^+ \rangle$ in Fig.~\ref{figure4}(a) and the average streamwise velocity profile shifted by the average streamwise slip velocity in wall units $\langle U^+ \rangle - U_s^+$ in Fig.~\ref{figure4}(b).
We consider $\langle U^+ \rangle$ and $\langle U^+ \rangle - U_s^+$ separately in order to validate the average streamwise and spanwise slip mechanisms introduced in \eqref{eq:shift} and their contribution to the drag reduction.

By comparing the curves for no-slip and SHS walls in Fig.~\ref{figure4}(a) for $y^+ \in [0, \, 5]$, the average streamwise velocity (in wall units) is shifted upwards by the average streamwise slip velocity (in wall units) inside the viscous sublayer, as discussed in \citet{turk14}.
Increasing the SHS period in wall units for a fixed gas fraction increases the average streamwise slip velocity at the SHS and decreases the drag.
By comparing curves for no-slip and SHS walls in Fig.~\ref{figure4}(b) for $y^+ \in [25, \, 300]$, the average streamwise velocity minus the average streamwise slip velocity (in wall units) is shifted downwards inside the log law \citep{turk14}.
Increasing the SHS period in wall units for a fixed gas fraction increases the average spanwise slip velocity at the SHS, allowing streamwise vortices to move closer to the SHS. 
This enhances turbulent momentum transfer close to the SHS and increases the drag.
The spanwise slip mechanism is less dominant than the streamwise slip mechanism, as the ridges are much longer in the streamwise direction than in the spanwise direction. 

\subsection{Comparison with laboratory experiments} \label{Comparison with experiments including surfactant}

\subsubsection{Turbulent model excluding surfactant}

\begin{figure}[t!]
    \centering
    \includegraphics[width=\columnwidth,trim={3cm 0 4cm 1.5cm},clip]{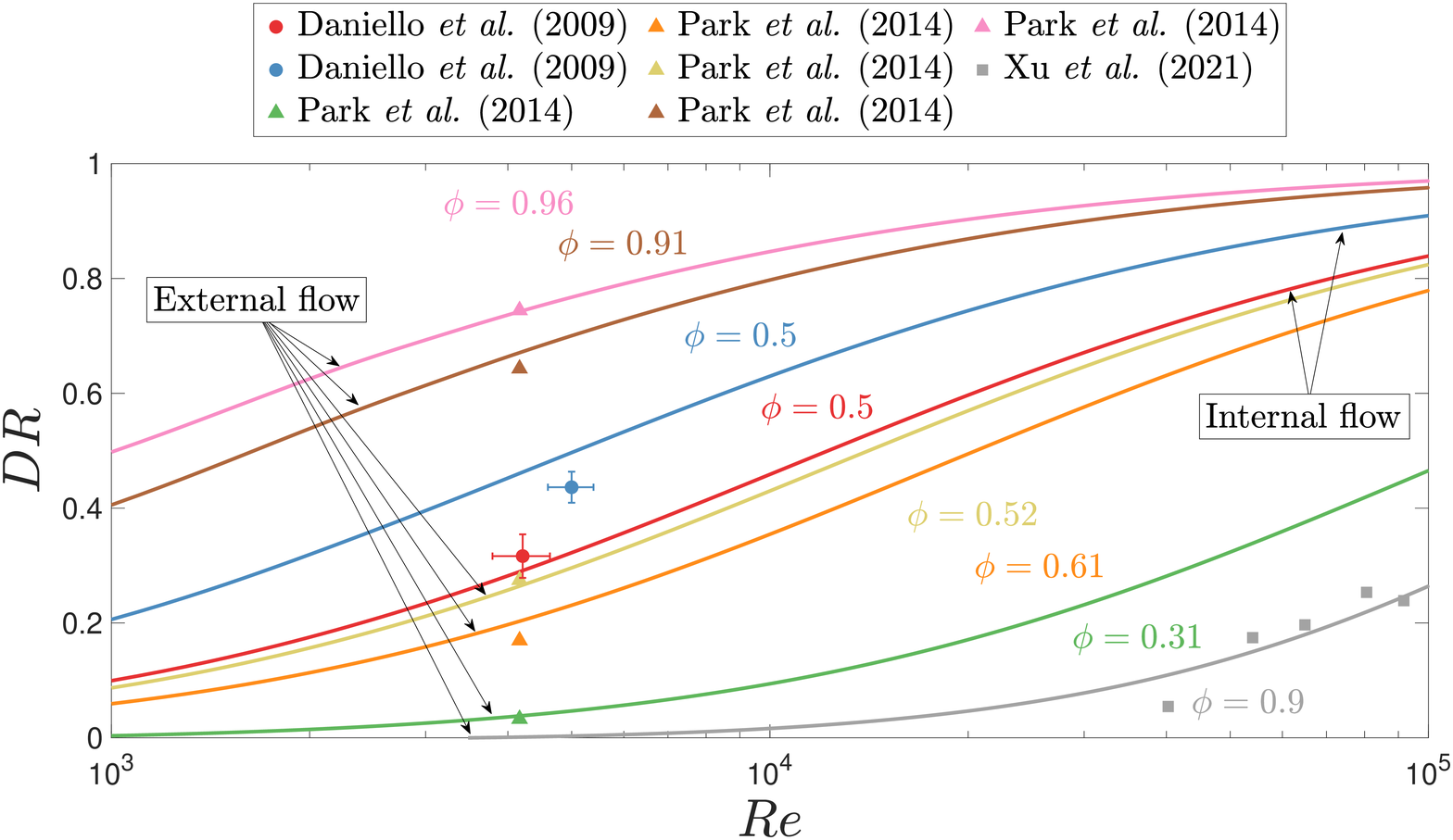} \\
	\caption{Comparison of our turbulent model predictions for the drag reduction (${DR}$) using (\ref{eq:dragreductiondefinition}, \ref{eq:equivilentflux2highre}) with experimental results in the literature for internal turbulent channel flows with SHS \citep{Daniello_etal_PF_2009} and external turbulent boundary layer flows with SHS \citep{Park2014, xuJFM21}, whilst varying the Reynolds number (${Re}$) and gas fraction ($\phi$), where the average Marangoni shear rate $\gamma_{Ma}=0$. 
	Filled symbols and solid lines represent turbulent experiments and theory.}
	\label{figure5}
\end{figure}

We compare the drag reduction predicted by our turbulent model using (\ref{eq:dragreductiondefinition}, \ref{eq:equivilentflux2highre}) (excluding surfactant effects, such that $\gamma_{Ma}=0$) with the available experimental data in the literature for turbulent flows over SHSs \citep{Daniello_etal_PF_2009, Park2014, xuJFM21}, as a function of the Reynolds number ${Re}\in[1000, 100000]$ and gas fraction $\phi = 0.31$ (green symbols and lines), $\phi = 0.5$ (blue and red), $\phi = 0.52$ (yellow), $\phi = 0.61$ (orange), $\phi = 0.9$ (grey), $\phi = 0.91$ (brown) and $\phi = 0.96$ (pink). 
We note that no surfactant was  added artificially in the  experiments above from  the literature. 
However, surfactants may  have been  present in these experiments in small amounts from contamination due to laboratory conditions and equipment \cite[e.g. microfluidic devices made of PDMS have been shown to lead to surfactant effects in][]{Hourlier-Fargette2018-tu}.
In contrast to Section \ref{ref:Comparison with direct numerical simulations excluding surfactant} and based on the information presented in these experimental studies, we cannot present results on the average streamwise slip length. 
Local quantities, such as $\lambda_x^+$, are much harder to measure than global quantities (i.e. ${DR}$) in experiments because of the small length scales associated with flows over SHSs.

We first discuss how the experimental configuration changes the turbulent drag reduction for flows over SHSs and then use this to explain the non-monotonicity of ${DR}$ with respect to $\phi$ in Figure \ref{figure5}.
Similar to the texture-resolving DNS results presented in Section \ref{ref:Comparison with direct numerical simulations excluding surfactant}, \citet{Daniello_etal_PF_2009} considered an internal flow configuration bounded by SHSs with streamwise ridges that are periodic in the spanwise direction with $\phi = 0.5$ (red and blue symbols and lines).
The experimental works of \citet{Park2014} and \citet{xuJFM21} consider turbulent flows over a test section with finite streamwise ridges that are periodic in the spanwise direction (pink, brown, yellow, orange, green and grey). 
The turbulent boundary layer thickness must first be obtained in order to evaluate the drag reduction using (\ref{eq:dragreductiondefinition}, \ref{eq:equivilentflux2highre}) (see Fig.~\ref{figure2}b).
A boundary layer originates from the leading edge of the channel in \cite{Park2014} and the plate in \citet{xuJFM21}, developing over approximately $\SI{45}{\centi\metre}$  and $\SI{1.1}{\metre}$, respectively, measured from the leading edge to the centre of the SHS test section. 
For the purpose of this study, we will assume that the turbulent boundary layer thickness $H = H(x)$ can be approximated by the classical result from turbulent boundary-layer theory \citep{schlichting2003boundary}, $H = 0.37 x / {Re}_x^{1/5}$, where ${Re}_x = U x/\nu$ is the boundary layer Reynolds number and $x$ is the distance from the leading edge to the centre of the test section, as done in \citet{xuJFM21}.
We now use the above boundary-layer approximation to highlight an important difference between  configurations with varying $H/P$ in external flows to explain the non-monotonicity of ${DR}$ with respect to $\phi$. 
The ratio $H / P$ varies significantly if we compare the experimental setup that generates the brown curve of \citet{Park2014} (where the distance from the leading edge to the centre of the test section is \SI{45}{\centi\metre}) and the experimental setup that generates the grey curve of \citet{xuJFM21} (\SI{1.1}{\metre}). 
This change in $H / P$ causes the drag reduction to be smaller in \citet{xuJFM21} even though the gas fraction $\phi = 0.9$ and texture period $P = \SI{50}{\micro\metre}$ do not change across the two experiments.

The model in (\ref{eq:dragreductiondefinition}, \ref{eq:equivilentflux2highre}) captures the increase in drag reduction with increasing gas fraction in the experiments of \citet{Park2014} (see the green, orange, yellow, brown and pink curves in Fig.~\ref{figure5}).
The orange data where $\phi = 0.61$ exhibit a smaller drag reduction than the yellow data where $\phi = 0.52$, as the texture period has decreased from $P = \SI{100}{\micro\metre}$ to $P = \SI{50}{\micro\metre}$ \citep{Park2014}, reducing the area of the liquid--gas interface at the SHS.
The same effect is noticed in the experimental data from \citet{Daniello_etal_PF_2009} by comparing results for $P = \SI{30}{\micro\metre}$ (red) and $P = \SI{50}{\micro\metre}$ (blue).
There is a significant spread in the original experimental data presented in \citet{Daniello_etal_PF_2009}, which could be due to a number of features of SHSs, such as the liquid--gas interface curvature, the gas subphase, loss of plastron and ridge misalignment \citep{park21}.
In Fig.~\ref{figure5} we show an ensemble average (error bars) of the drag-reduction data extracted from \citet{Daniello_etal_PF_2009} over all Reynolds numbers in order to simplify the comparison between these data and the other experiments.

\subsubsection{Turbulent model including surfactant}

\begin{figure*}[t!]
	\centering
	(a) \hfill (b) \hfill (c) \hfill \hfill \hfill \\ \includegraphics[width=.33\textwidth,trim={.5cm 0 1cm 0},clip]{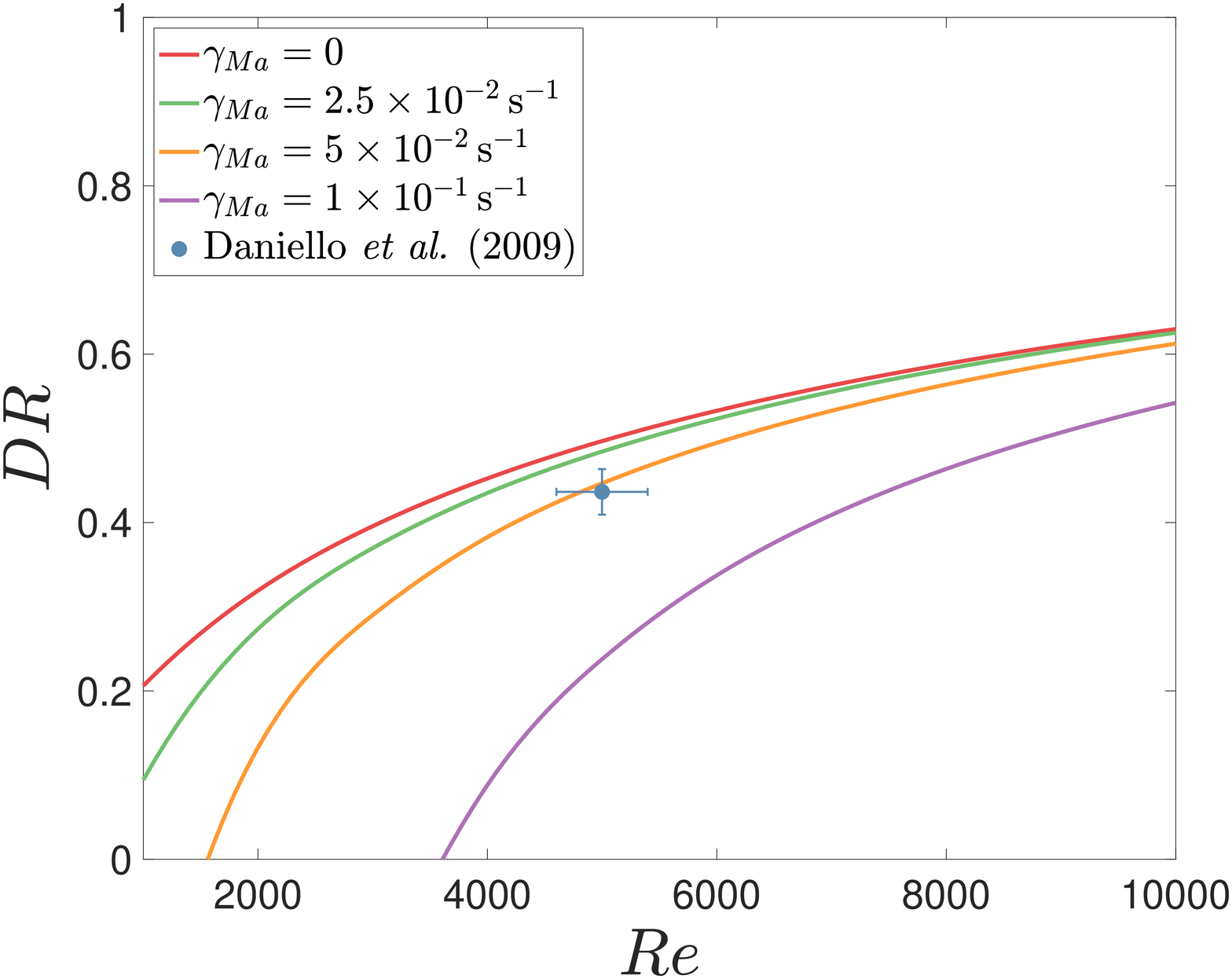}\includegraphics[width=.33\textwidth,trim={.5cm 0 1cm 0},clip]{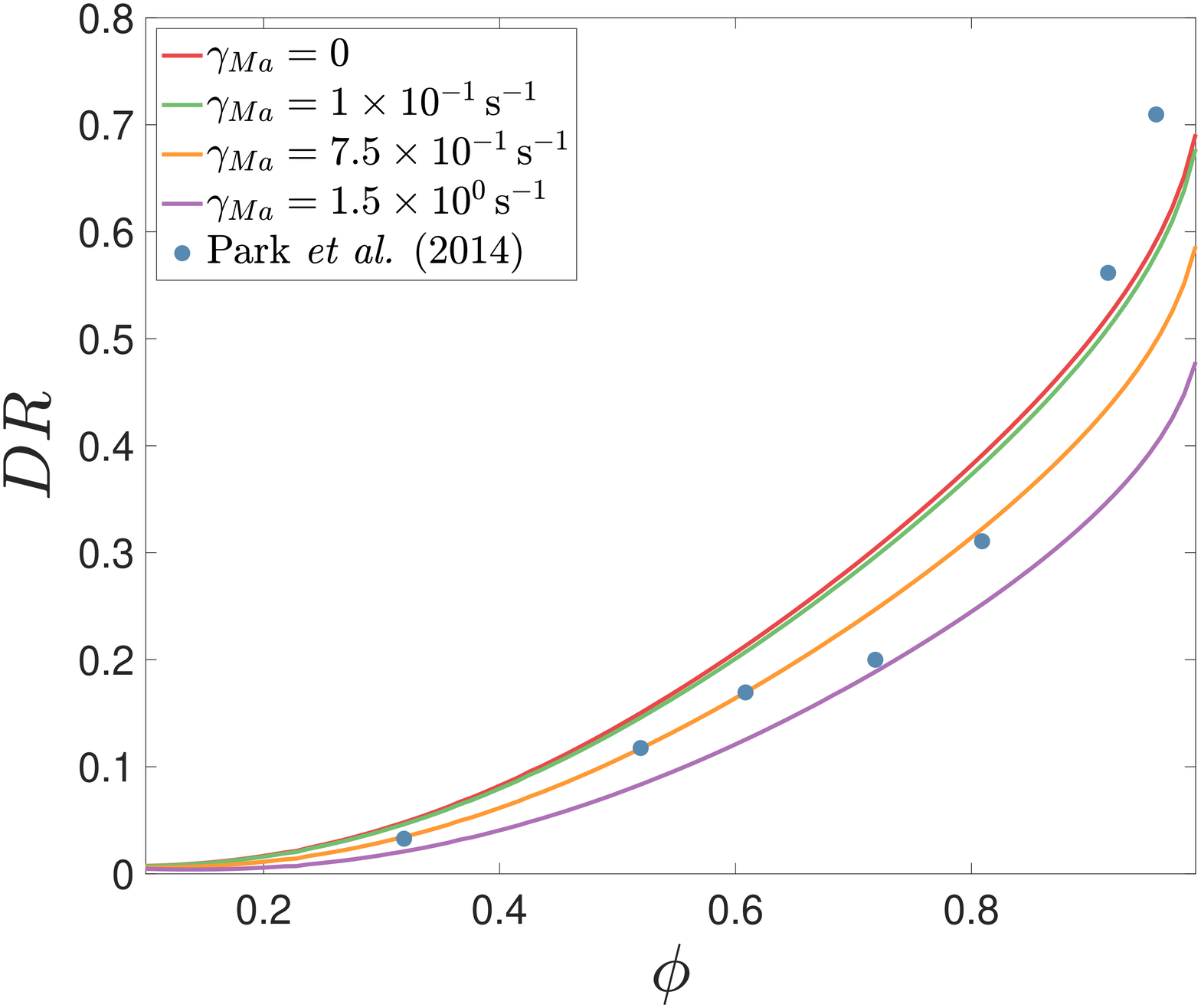}\includegraphics[width=.33\textwidth,trim={.5cm 0 1cm 0},clip]{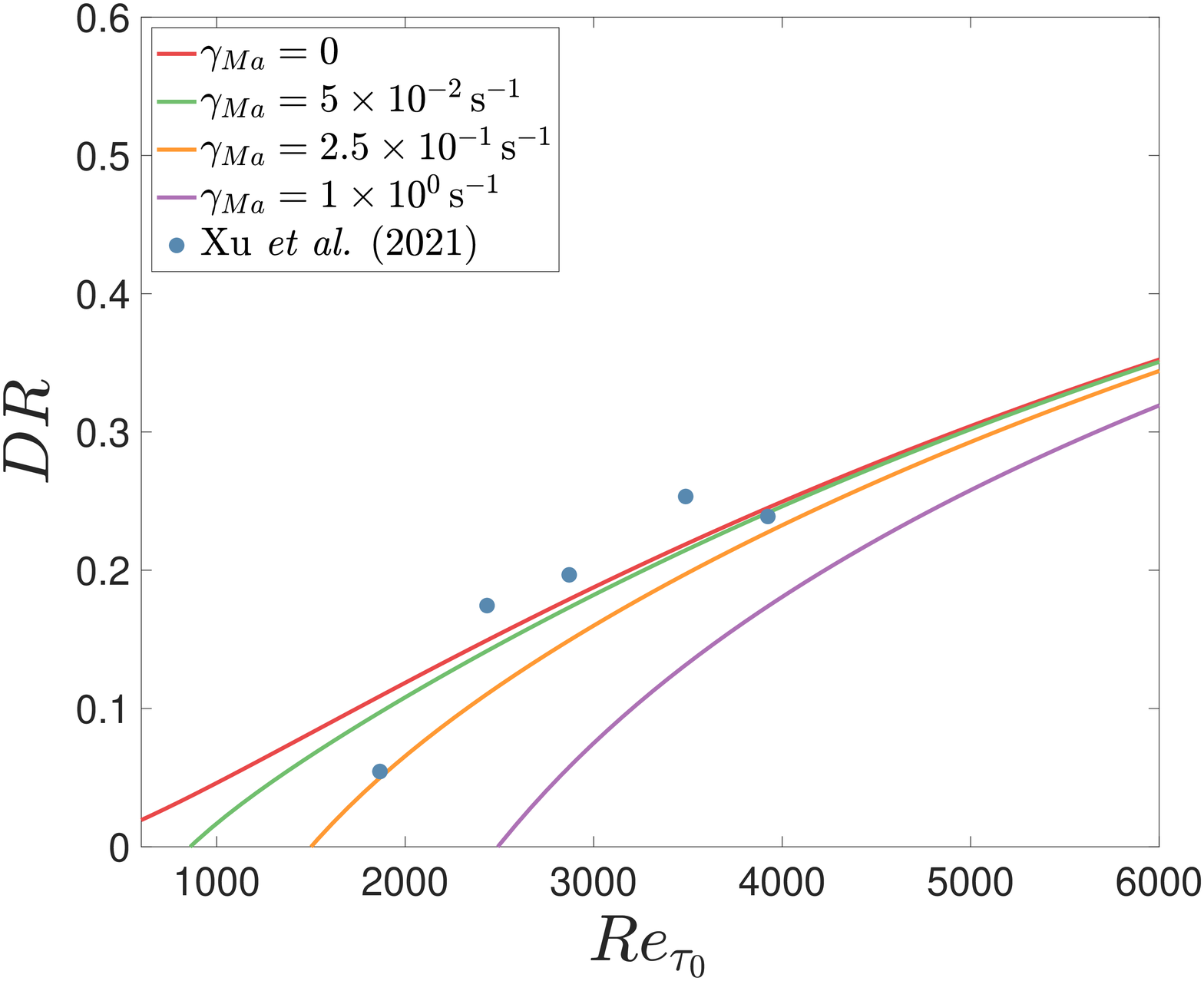} \\
    \caption{Comparison of our turbulent model predictions (solid lines) for the drag reduction ${DR}$ using (\ref{eq:dragreductiondefinition}, \ref{eq:equivilentflux2highre}) at various Marangoni shear rates $\gamma_{Ma}$ (colours) with  experimental data (symbols) from the literature: (a) \citet{Daniello_etal_PF_2009} for varying Reynolds number (${Re}$); (b) \citet{Park2014} for varying gas fraction ($\phi$); (c) \citet{xuJFM21} for varying no-slip friction Reynolds number (${Re}_{\tau_0})$. 
    }
	\label{figure6}
\end{figure*}

We now investigate the potential effect of surfactants on the drag reduction of flows over SHSs in experiments reported in the literature. 
Since surfactants have generally not been added artificially in experiments in the literature, the concentration level and type of any potential surfactant present in the experiments is unknown. 
In Fig.~\ref{figure6}, we compute theoretical predictions from our model using (\ref{eq:dragreductiondefinition}, \ref{eq:equivilentflux2highre}) for varying  average Marangoni shear rate $\gamma_{Ma}$, thereby simulating different surfactant conditions. 
We compare our turbulent model predictions inclusive of surfactant with  available experimental data from: \citet{Daniello_etal_PF_2009} in panel (a), \citet{Park2014} in panel (b) and \citet{xuJFM21} in panel (c), as a function of the bulk Reynolds number ${Re}\in[1000,\, 10000]$, gas fraction $\phi \in[0.1, \, 0.9]$ and friction Reynolds number ${Re}_{\tau_0} \in[0,\, 8000]$ respectively. 
Our theoretical model shows that the effect of surfactant is stronger at lower Reynolds numbers, where the average Marangoni shear rate $\gamma_{Ma}$ at the liquid--gas interface is large compared to the average wall shear stress. 
This can be seen in panels (a) and (c) where $DR$ decreases more rapidly with increasing $\gamma_{Ma}$ (coloured solid lines) at smaller $Re$ and $Re_{\tau_0}$, respectively. 
When ${DR} = 0$, the surfactant is strong enough to immobilise the liquid--gas interface, such that the mean streamwise slip velocity is zero.
For those configurations where the liquid--gas interface becomes immobilised at a fixed finite Reynolds number, the interface remains immobilised at all Reynolds numbers that are smaller than this value (see, for example, the purple curve in Fig.~\ref{figure6}(a), where the interface is immobilised for all ${Re} \lessapprox 3500$).
As the gas fraction of the SHS decreases in the limit $\phi \rightarrow 0$, there is no interface for the surfactant to adsorb to, and therefore, the curves for different $\gamma_{Ma}$ collapse and ${DR} \rightarrow 0$.

Overall, we find that theoretically the inclusion of surfactant effects in our turbulent model can clearly impair drag reduction. 
Nevertheless, the limited amount of experimental data  is not sufficient to confirm or infirm the impact of surfactants in the experiments we have analysed.
The experimental data plotted in Fig.~\ref{figure5} and Fig.~\ref{figure6} do not strongly deviate from the model assuming $\gamma_{Ma}=0$, thus suggesting weak or negligible surfactant impact in the laboratory experiments that we have analysed from \citet{Daniello_etal_PF_2009}, \citet{Park2014} and \citet{xuJFM21}. 
In Table \ref{tab:my_label}, surfactant effects are quantified via the root mean squared (RMS) error $\epsilon_{\text{RMS}}$ which compares the drag reduction predicted by our model ${DR}_{\text{Model}}$ to the drag reduction predicted by experimental data ${DR}_{\text{Data}}$.
We see that for the experimental data in \citet{Daniello_etal_PF_2009}, \citet{Park2014} and \citet{xuJFM21}, the predictions for weak surfactant effects with a small non-zero Marangoni shear rate $\gamma_{Ma}$ give rise to a smaller RMS error than those for a clean channel where $\gamma_{Ma}=0$. 
Conversely, the predictions for moderate or strong surfactant effects with a larger $\gamma_{Ma}$ have a greater $\epsilon_{\text{RMS}}$ than those channels with $\gamma_{Ma}=0$.
The limited data and lack of experiments including surfactant make these experimental results difficult to interpret.  
One would expect the effect of surfactants to be more prominent in fieldwork rather than in a laboratory setting, where the water is relatively clean. 
We discuss our model predictions when surfactant concentrations and ridge lengths are characteristic of marine applications in Section \ref{Model predictions}.
More experiments that vary the surfactant concentration are therefore required to infer whether surfactants are important in turbulent applications. 
As previously mentioned, several additional features of flows over SHSs could be involved and cause the changes in drag; e.g. liquid--gas interface curvature, the gas subphase, loss of plastron or ridge misalignment \citep{park21}. 

\renewcommand{\arraystretch}{1.25}
\begin{table}[t!]
\resizebox{\columnwidth}{!}{%
    \centering
    \begin{tabular}{c|cccc}
    \hline
    \hline
    \multicolumn{5}{c}{Daniello \textit{et al.} (2009)} \\
    \hline
    $\gamma_{Ma}$ (s$^{-1}$) & $ 0$ & $2.5\times 10^{-2}$ & $5\times 10^{-2}$ & $1\times 10^{-1}$ \\
    $\epsilon_{\text{RMS}}$ & 0.0173 & 0.0023 & 0.0001 & 0.0173 \\
    \hline
    \hline
    \multicolumn{5}{c}{Park \textit{et al.} (2014)} \\
    \hline
    $\gamma_{Ma}$ (s$^{-1}$) & $ 0$ & $1\times 10^{-1}$ & $7.5\times 10^{-1}$ & $1.5\times 10^{0}$ \\
    $\epsilon_{\text{RMS}}$ & 0.0052 & 0.0052 & 0.0090 & 0.0210 \\
    \hline
    \hline
    \multicolumn{5}{c}{Xu \textit{et al.} (2021)} \\
    \hline
    $\gamma_{Ma}$ (s$^{-1}$) & $ 0$ & $5\times 10^{-2}$ & $2.5\times 10^{-1}$ & $1\times 10^{0}$ \\
    $\epsilon_{\text{RMS}}$ & 0.0010 & 0.0009 & 0.0019 & 0.0215 \\
    \hline
    \hline
    \end{tabular}
}
    \caption{The RMS error of our model, $\epsilon_{\text{RMS}}$, comparing the drag reduction predicted by our model (${DR}_{\text{Model}}$) using (\ref{eq:dragreductiondefinition}, \ref{eq:equivilentflux2highre}) to the drag reduction predicted by laboratory experimental data (${DR}_{\text{Data}}$), considering experimental results in internal \citep{Daniello_etal_PF_2009} and external flows \citep{Park2014, xuJFM21} from the literature, for different average Marangoni shear rates ($\gamma_{Ma}$).}
    \label{tab:my_label}
\end{table}

\subsection{Model predictions for marine applications} \label{Model predictions}

\begin{figure}[t!]
    \centering
    \includegraphics[width=\columnwidth,trim={3.3cm 0.3cm 4.2cm .6cm},clip]{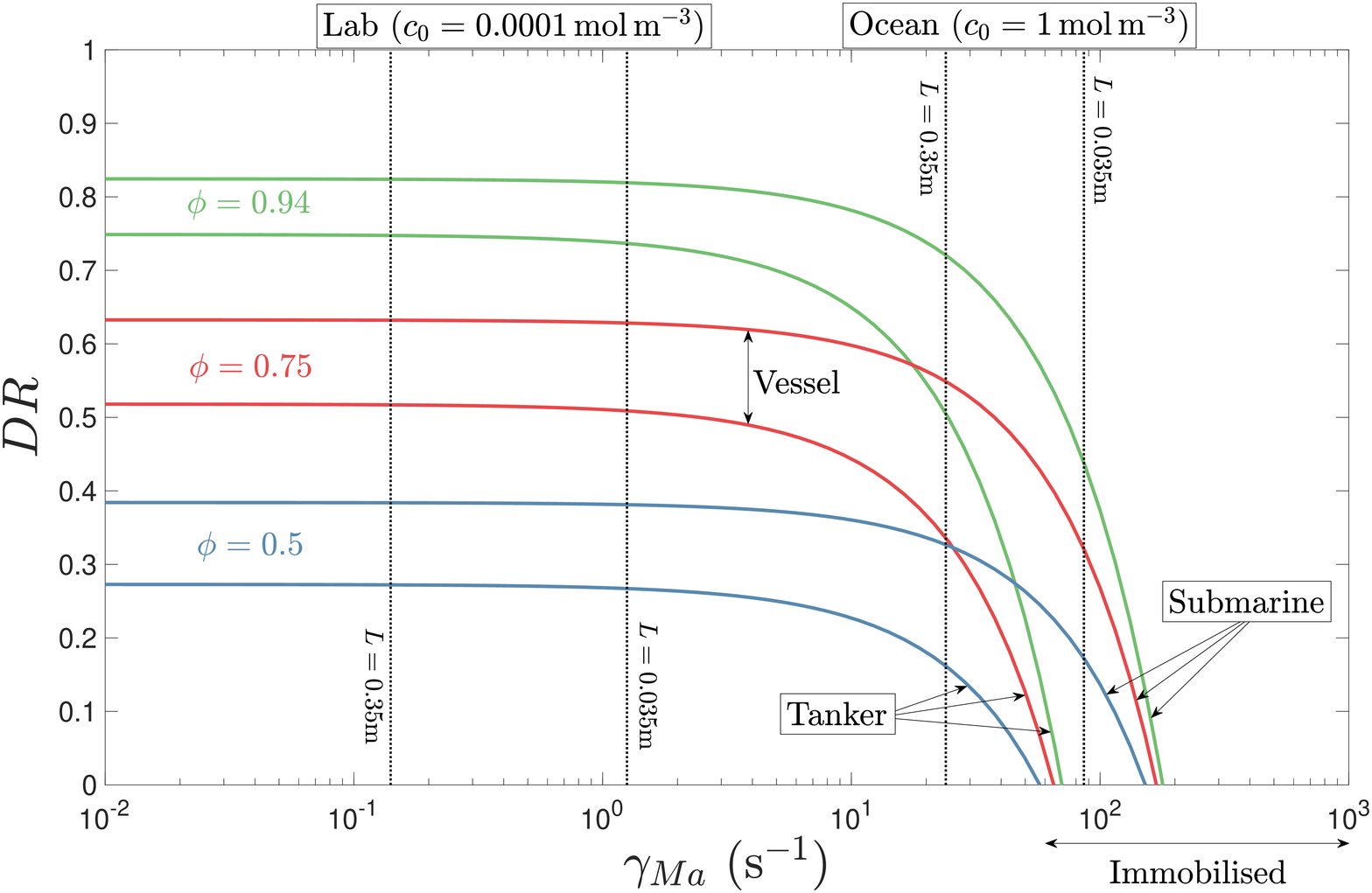} \\
	\caption{Turbulent model predictions for the drag reduction (${DR}$) using (\ref{eq:dragreductiondefinition}, \ref{eq:equivilentflux2highre}) in laboratory (where the bulk concentration $c_0 = \SI{1e-4}{\mole\per\metre\cubed}$ and the streamwise ridge length $\SI{0.035}{\metre} \leq L \leq \SI{0.35}{\metre}$, indicated by the leftmost vertical black dotted lines) and ocean environments (where $c_0 = \SI{1}{\mole\per\metre\cubed}$ and $\SI{0.035}{\metre} \leq L \leq \SI{0.35}{\metre}$, indicated by the rightmost vertical black dotted lines), whilst varying the average Marangoni shear rate ($\gamma_{Ma}$), for different gas fractions ($\phi$) and applications detailed in Table \ref{tab:Re}. 
    For each gas fraction, the upper bound gives ${DR}$ for a submarine and the lower bound gives ${DR}$ for a tanker.}
	\label{figure7}
\end{figure}

We finally investigate how the drag reduction varies with respect to the average Marangoni shear rate $\gamma_{Ma}$, which arises due to surfactant accumulation (with background concentration $c_0$) at the downstream stagnation point of the long but finite streamwise ridges (with length $L$) that are periodic in the spanwise direction.
We compute predictions for ${DR}$ across a range of $\gamma_{Ma}$ in Fig.~\ref{figure7}, using the model in (\ref{eq:dragreductiondefinition}, \ref{eq:equivilentflux2highre}) with $\phi = 0.5$ (blue curves), $\phi = 0.75$ (red) and $\phi = 0.94$ (green), for a range of length and velocity scales characteristic of marine applications.
In Table \ref{tab:Re}, we present these typical length and velocity scales that are characteristic of marine applications, such as a tanker or submarine. 
The data in Table \ref{tab:Re} is used to calculate the bulk Reynolds number ${Re}$ and turbulent boundary-layer thickness $H$, which is approximated using $H = 0.37 x / {Re}_x^{1/5}$ \citep{schlichting2003boundary}, for the equivalent no-slip flow. 
The approximate turbulent boundary-layer thickness is evaluated at the streamwise mid-point of the marine vessels considered in Table \ref{tab:Re}, such that it lies within the range $\SI{0.15}{\metre} \leq H \leq \SI{0.35}{\metre}$.
We choose a value for the SHS texture period $P$ based on those SHSs that have been reported to maintain a stable Cassie--Baxter state in experiments in the literature \citep{Daniello_etal_PF_2009, jung10, Park2014, Woolford2009, xuJFM21}: $\SI{100}{\micro\metre} \leq P \leq \SI{200}{\micro\metre}$, i.e. we take $P = \SI{150}{\micro\metre}$. 

\begin{table*}[t!]
    \centering
    \begin{tabular}{l|cccccccc}
    \hline
    \hline
        Vessel & Length & Speed & $Re$ & $H$ & $P$ & $L$ & $c_0$ \\
        & \SI{}{\metre} & \SI{}{\metre \per \second} & - & \SI{}{\metre} & \SI{}{\metre} & \SI{}{\metre} & \SI{}{\mole \per \metre \cubed} \\
        \hline
        Tanker & 400 & 8.5  & $4.3 \times 10^8$  & 0.35 & $1.5 \times 10^{-4}$ & [0.035, 0.35] & [0.0001, 1] \\
        Submarine & 150 & 13 & $2.4 \times 10^8$  & 0.15 & $1.5 \times 10^{-4}$ & [0.035, 0.35] & [0.0001, 1] \\
            \hline
    \hline
    \end{tabular}
    \caption{Table showing the typical length, speed, bulk Reynolds number (based on the speed and length of the vessel and the kinematic viscosity of water), boundary-layer thickness, pitch \cite[based on][]{Daniello_etal_PF_2009, jung10, Park2014, Woolford2009, xuJFM21}, streamwise ridge length \citep{xuJFM21} and background concentration \citep{frossard2019properties, temprano2021slip}, for various marine vessels.
    These length, velocity and concentration scales are used to evaluate the drag reduction for a tanker and submarine in Fig.~\ref{figure7}.}
    \label{tab:Re}
\end{table*}

We also estimate the average Marangoni shear rate $\gamma_{Ma}$ in lab and ocean environments using the theory outlined in \citet{landel19}, with the characteristic velocities $U$, boundary layer thicknesses $H$, streamwise ridge lengths $L$ and background concentrations $c_0$ that are summarised in Table \ref{tab:Re}.
The scaling theory derived in \citet{landel19} approximates the surfactant dynamics using a linear equation of state and adsorption--desorption kinetics (see \ref{sec:sample:appendix0}). 
In order to use this model, we have assumed that the spanwise variations in the velocity and concentration fields are negligible compared to the streamwise variation. 
Indeed, in the experiments conducted by \citet{xuJFM21}, the gas fraction is large, $\phi = 0.9$, and therefore, we would expect three-dimensional effects to be small. 
The validity of the above assumptions in turbulent flows over SHSs with surfactant is left for future study.
We base the streamwise ridge length of the SHS on the configuration in \citet{xuJFM21} where a stable liquid--gas interface was mostly maintained; these experiments took place for ${Re} \in [2.3 \times 10^{6}, \, 1.12 \times 10^7]$, which is closest to the marine applications that we investigate in this study. 
First, we let the total length of the streamwise ridges be $L=\SI{0.035}{\metre}$ and the length of the solid region between ridges to be \SI{30}{\micro\metre}.
We then allow for the possibility of longer ridges than those considered in \citet{xuJFM21}, i.e. $L=\SI{0.35}{\metre}$, primarily to demonstrate how $\gamma_{Ma}$ depends on $L$. 
We plot the average Marangoni shear rate that we estimate to be characteristic of laboratory environments $\gamma_{Ma} = \SI{0.14}{\per\second}$ when $L = \SI{0.35}{\metre}$ and $\gamma_{Ma} =\SI{1.25}{\per\second}$ when $L = \SI{0.035}{\metre}$ (leftmost vertical black dashed lines), where we expect surfactant concentrations to be low, i.e. $c_0 = \SI{1e-4}{\mole\per\metre\cubed}$, as estimated in lab conditions by \citet{temprano2021slip}. 
We also plot the average Marangoni shear rate that we assume to be characteristic of ocean environments $\gamma_{Ma} = \SI{23.97}{\per\second}$ when $L = \SI{0.35}{\metre}$ and $\gamma_{Ma} = \SI{86.08}{\per\second}$ when $L = \SI{0.035}{\metre}$  (rightmost vertical black dashed lines) where the surfactant concentration can be much higher, i.e. $c_0 = \SI{1}{\mole\per\metre\cubed}$, as measured in ocean conditions by \citet{frossard2019properties}.

In Fig.~\ref{figure7}, we find that the surfactant concentrations that are characteristic of clean laboratory conditions are not high enough to develop an appreciable surfactant gradient and increase the drag for flows with this particular SHS geometry. 
Hence, our model predicts that surfactant effects are weak in this regime.
The surface velocity is large in turbulent flows, which means that the shear rate of the SHS flow is greater than the shear rate due to surfactant, and the liquid--gas interface is effectively shear-free.
However, the higher surfactant concentrations that are present for marine applications in the ocean mean that the shear rate due to surfactant increases, and therefore, a surfactant gradient might develop at the liquid--gas interface that generates an appreciable increase in the drag for flows with this particular SHS geometry. 
Hence, our model predicts that surfactant effects are moderate to strong in this regime.
For example, in Fig.~\ref{figure7}, surfactant effects are strong and the interface is immobilised (i.e. ${DR} = 0$) when the background concentration is larger than a threshold we estimate at $c_0 \gtrsim \SI{1}{\mole \per \metre \cubed}$. 
Immobilisation occurs for a smaller $\gamma_{Ma}$ for tanker applications when compared to submarine applications, as the characteristic velocities are typically slower (see Table \ref{tab:Re}).

\section{Conclusions}
\label{sec:Conclusions}

Motivated by recent developments that demonstrate the importance of surfactants in laminar flows over SHSs \citep{Kim2012-iw, Bolognesi2014-vw, peaudecerf17, Song2018-uw, landel19, temprano2021slip}, we have proposed a model for turbulent flow over SHSs with long but finite streamwise ridges that are periodic in the spanwise direction, including surfactant effects, based on the shifted-log-law theory applied to SHSs by \citet{Fukagata2006-hk}. 
We consider both internal and external flows over SHSs, in order to compare with the wide range of numerical \citep{Park2013, turk14, Rastegari2015, egan2021direct} and experimental \citep{Daniello_etal_PF_2009, Park2014, xuJFM21} data in the literature and predict the drag reduction for marine applications.
The turbulent model assumes that the viscous sublayer thickness is much larger than the SHS texture period $P$, and therefore, that the SHS texture affects the turbulent bulk flow via the average streamwise and spanwise slip length.
Our model employs an empirical relationship for the saturation of the log-law shift due to the average spanwise slip length based on riblet theory \citep{luchini91, ibrahim21} and applied to SHSs by \citet{busse12}.
We close the model using laminar solutions due to \citet{Philip_ZAMP_1972a}, where we extend the solutions in \citet{Philip_ZAMP_1972a} to include surfactant effects.
This provides us with a fully predictive relationship for the turbulent drag reduction, which we can use to relate the turbulent drag reduction to the geometry of the SHS, the flow, the fluid and the properties of the surfactant, using a laminar scaling theory outlined in \citet{landel19}.

We compare our model predictions with direct numerical simulations (DNS), where there is good agreement in the drag reduction for small $P^+$ (in wall units $+$), i.e. when the viscous sublayer is thick compared to the period of the SHS (Fig.~\ref{figure3}).
The model captures the dependence of the drag reduction on the cross-plane geometry of the SHS, i.e. the gas fraction $\phi$, texture wavelength $P$ and the wall-normal height $H$, where the streamwise and spanwise slip mechanisms that give rise to the drag reduction can be examined using the flow field (Fig.~\ref{figure4}).
The agreement between the model and DNS holds for $P^+ \lessapprox 50$ until we transition into a different regime that is dominated by turbulence for $P^+ \gtrapprox 50$, where the drag reduction from the DNS asymptotically approaches the gas fraction for $P^+ \rightarrow \infty$, as also discussed by \citep{Daniello_etal_PF_2009, Rothstein:2010im, park21}.
We calculate the streamwise slip length that corresponds to this empirical asymptote to improve model predictions at large $P^+$.
We also compare our model predictions with experimental data in nominally clean (i.e. where no surfactants were added artificially) laboratory settings (Fig.~\ref{figure5}), which allows us to investigate any potential contaminant surfactant effects in turbulent flows over SHSs. 
The theory demonstrates that the presence of surfactant is detrimental to drag reduction, where greater increases in drag are seen at smaller Reynolds numbers.
By comparing the surfactant-inclusive model and the laboratory experimental data found in the literature, our model shows that surfactants did not affect significantly the drag reduction performance of the SHSs studied in laboratory conditions (Fig.~\ref{figure6}), as expected from clean experimental conditions. 
For shorter gratings, which are necessary at high speeds to maintain a stable liquid--gas interface \cite[see e.g. $L=\SI{0.035}{\metre}$ in][]{xuJFM21}, and higher surfactant concentrations which have been measured in the ocean \citep{frossard2019properties}, our model predicts that surfactant can become important again for  velocities and length scales characteristic of marine applications (Fig.~\ref{figure7}).
Both DNS including surfactant dynamics and experimental studies with surfactant concentrations that are typical of ocean environments  are required to further disentangle the effect of surfactants in turbulent flows over SHSs. 

\section*{Declaration of Competing Interest}

The authors declare that they have no known competing financial interests or personal relationships that could have appeared to
influence the work reported in this paper.

\section*{Acknowledgements}

We acknowledge support from CBET--EPSRC (EPSRC Ref. EP/T030739/1, NSF \#2054894), as well as partial support from ARO MURI W911NF-17-1-0306.

\appendix

\section{Scaling theory for the average Marangoni shear rate}
\label{sec:sample:appendix0}

For completeness, we outline one of the main results from the scaling theory derived in \citet{landel19}, so that we can discuss the dependence of the average Marangoni shear rate $\gamma_{Ma}$ on the wall-normal height $H$, streamwise length of the ridges $L$, characteristic velocity $U$ and background bulk surfactant concentration $c_0$.
\citet{landel19} consider a steady, two-dimensional pressure-driven channel flow bounded by a single SHS, which is contaminated with a small concentration of surfactant. 
They linearise the equation of state and adsorption--desorption kinetics and perform a scaling analysis on the resulting governing equations.  
By solving for the two-dimensional velocity field using dual series techniques and combining this with the scaling analysis results, they find that 
\begin{equation} \label{eq:gam_ma}
    \gamma_{Ma} = \frac{a_1 k {Ma} F_0 U}{\displaystyle H \left(\frac{1}{Pe_I} + \frac{a_2 L^2 {Bi}\chi}{\chi + {Bi} {Pe} \delta} + a_1 k {Ma} F_0\right)},
\end{equation}
where $a_1 \approx 2.3$ and $a_2 \approx 0.32$ are empirical parameters that are fitted using simulations, $k = k_a c_0 /k_d$ is the bulk concentration, ${Ma} = n R T \Gamma_m /\mu/U$ is the Marangoni number, $F_0$ is the interfacial velocity of the clean flow \cite[see][for more details]{landel19}, ${Pe}_I = H U /D_I$ is the interfacial P\'{e}clet number, ${Bi}=k_d H /U$ is the Biot number, $\chi = k_d H / k_a / \Gamma_m$ is the kinetics number, ${Pe} = H U / D$ is the bulk P\'{e}clet number and $\delta \approx 1.68 (L/H) (1 + 0.05 (L/H)^2 Pe)^{-1/3}$ is the typical thickness of the diffusive layer of bulk surfactant.
The dimensional surfactant parameters that are used to calculate the above non-dimensional numbers and generate the results in Fig. 7 are given in Table 3.  
From \eqref{eq:gam_ma}, we observe that as the bulk surfactant concentration increases, the dimensionless group $a_1 k {Ma} F_0$ increases and the average Marangoni shear rate increases. 
Conversely, we observe that as the streamwise ridge length increases, the dimensionless group $a_2 L^2 {Bi} \chi / (\chi + {Bi} {Pe} \delta)$ increases and the average Marangoni shear rate decreases. 

\begin{table}[t!]
    \centering
\resizebox{\columnwidth}{!}{%
    \begin{tabular}{c|c|c|c}
    \hline
    \hline
        Quantity & Symbol & Units & Value \\
        \hline
Adsorption rate & $k_a$ & \SI{}{\metre \cubed \per \mole \per \second} & $89.5$ \\
Desorption rate & $k_d$ & \SI{}{\per \second} & $500$ \\
Salinity parameter & $n$ & - & $2$ \\
Ideal gas constant & $R$ & \SI{}{\joule \per \mole \per \kelvin} & $8.31$ \\
Temperature & $T$ & \SI{}{\kelvin} & $296$ \\
Packing concentration & $\Gamma_m$ & \SI{}{\mole \per \metre \squared} & $3.9 \times 10^{-6}$ \\
Dynamic viscosity & $\mu$ & \SI{}{\kilogram \per \metre \per \second} & $8.9 \times 10^{-4}$ \\
Surface diffusivity & $D_I$ & \SI{}{\metre \squared \per \second} & $7 \times 10^{-10}$ \\
Bulk diffusivity & $D$ &  \SI{}{\metre \squared \per \second} & $7 \times 10^{-10}$ \\
    \hline
    \hline
    \end{tabular}
}
    \caption{Parameters appearing in the scaling theory for the average Marangoni shear rate \eqref{eq:gam_ma} from \citet{landel19} alongside their value used in the model prediction for marine applications in Fig. \ref{figure7}. }
    \label{tab:parameters}
\end{table}

\section{Converting direct numerical simulation data}
\label{sec:sample:appendix2}

In general, studies in the literature reporting  direct numerical simulations of SHS flows similar to our problem (see Fig.~\ref{figure3}) provide results only for the drag reduction ${DR}$ or the added flux $\Delta \overline{U}/\overline{U}_0$, but not both \citep{Park2013, turk14, Rastegari2015, egan2021direct}. 
The quantities ${DR}$ and $\Delta \overline{U}/\overline{U}_0$, defined in \eqref{eq:dragreductiondefinition} and  \eqref{eq:addedfluxdefinition} respectively, are two independent measurements of the performance of the SHS flow. 
Studies providing ${DR}$ were performed under the CFR condition \citep{Park2013, Rastegari2015}, whilst studies providing $\Delta \overline{U}/\overline{U}_0$ were performed under the CPG condition \citep{turk14, egan2021direct}.
In order to  compare  the numerical results for ${DR}$ with the largest data set from the literature, we have converted the results given for $\Delta \overline{U}/\overline{U}_0$ into ${DR}$. 
In the following, we describe our procedure to convert data for $\Delta \overline{U}/\overline{U}_0$ into data for ${DR}$. 
To minimise the conversion error, the procedure uses the log law for the no-slip flows \citep{Pope_CUP_2000}, whilst using the original published data for the SHS flows.


We convert the data for $\Delta \overline{U}/\overline{U}_0$, obtained under the CPG condition, into data for ${DR}$.  
In these simulations \citep{turk14, egan2021direct} the input parameters  include the prescribed stress $\tau=\tau_0$ and the wall-normal height $H$. 
To compute ${DR}$, we find the no-slip wall shear stress or friction Reynolds number obtained at the same bulk average velocity, i.e. $\tau_0(\overline{U})$ or $Re_{\tau_0}(\overline{U})$. 
If not given, we first seek $\overline{U}_0$ from \eqref{eq:Jintegral2}, knowing $Re_{\tau_0}$ and $H$. 
Then, we can obtain $\overline{U}$ from $\Delta \overline{U}/\overline{U}_0$ through \eqref{eq:addedfluxdefinition}. Finally, $\tau_0(\overline{U})$ or $Re_{\tau_0}(\overline{U})$ can be obtained using \eqref{eq:Jintegral2}, using $\overline{U}$ and $H$. 
The drag reduction is then computed as ${DR}=1-\tau(\overline{U})/\tau_0(\overline{U})$.

We note that, as long as the simulated no-slip flows are well resolved numerically, the  conversion procedure above should have a relatively small error as it only requires the use of classical log-law theory. The classical log-law theory should closely model the simulated no-slip flows in all the studies from which we have used data \citep{Park2013, turk14, Rastegari2015, egan2021direct}.

\section{Laminar streamwise velocity field including surfactant effects}
\label{sec:sample:appendix}

The laminar streamwise velocity field including surfactant effects can be found by solving the incompressible Stokes equation for a linear shear flow in a semi-infinite domain with free-stream shear-rate $\tau$.
The flow is assumed steady and homogeneous in the streamwise direction with a negligible pressure gradient.
The streamwise velocity is given by Laplace's equation
\begin{equation} \label{eq:lap}
    \frac{\partial^2 U}{\partial y^2} + \frac{\partial^2 U}{\partial z^2} = 0.
\end{equation}
The wall-normal and spanwise velocities are negligible as the streamwise length of the ridges is much larger than the cross-plane length scales (see Section \ref{sec:Formulation}). 
We solve \eqref{eq:lap} subject to a shear-rate condition at the liquid--gas interface \cite[which is derived from the linearised streamwise component of the tangential stress balance, where full details are given in][]{temprano2021slip}
\begin{equation} 
    \frac{\partial U}{\partial y}(y=0, \, 0\leq z \leq\phi) = \gamma_{Ma},
\end{equation}
no-slip conditions at the solid wall
\begin{equation} 
    U(y=0, \, \phi \leq z \leq P) = 0,
\end{equation}
symmetry conditions 
\begin{equation} 
    \frac{\partial U}{\partial z}(y,\, z = 0) = \frac{\partial U}{\partial z}(y,\, z = P) = 0,
\end{equation}
and a free stream shear rate
\begin{equation} 
    \lim_{y \rightarrow \infty} \frac{\partial U}{\partial y} = \frac{\tau}{\mu}. 
\end{equation}

Utilising superposition, we decompose the streamwise velocity field into one- and two-dimensional components. 
We can then solve for the two-dimensional component, using superposition to modify the conformal mapping solution due to \citet{Philip_ZAMP_1972a} to include surfactant effects through $\gamma_{Ma}$. 
Together, we have that
\begin{equation} \label{eq:sol}
    U = \frac{\tau y}{\mu} + \Im\left(\frac{P}{\pi}\left(\frac{\tau}{\mu} - \gamma_{Ma}\right) \arccos\left(\frac{\cos\left(\displaystyle \frac{\pi \theta }{P} \right)}{\cos\left(\displaystyle \frac{\pi \phi }{ 2}\right)}\right) - \theta \right),
\end{equation}
where $\theta = z + y \sqrt{-1}$, $i^2 = -1$ and $\Im(\cdot)$ denotes the imaginary part. 
Taking the limit as $y\rightarrow\infty$ of the difference between one and two-dimensional components, i.e. $U-\tau y/\mu$, we can evaluate the average streamwise slip length, $\lambda_x$, as \eqref{eq:laminarsliplengthx}.
This can then be used to evaluate the turbulent drag reduction, using the methodology outlined in Section \ref{sec:Model}.
As $\gamma_{Ma}\rightarrow 0$, we recover the original solution due to \cite{Philip_ZAMP_1972a} for a shear-free liquid--gas interface.
As $\gamma_{Ma} \rightarrow \tau / \mu$, the liquid-gas interface is immobilised and $U = \tau y / \mu$.

\bibliographystyle{elsarticle-harv} 
\bibliography{cas-refs}





\end{document}